\documentclass[utf8]{FrontiersinHarvard} 
\usepackage{url,hyperref,subcaption}
\usepackage{amsmath}

\def\NH2{$N\rm{(H_2)}$}

\def\SigmaH2{$\Sigma $(${\rm H_2}$)}

\def\cyclic{{\it c-}C$_3$H$_2$}
\def\CIII{HC$_3$N}
\def\CV{HC$_5$N}
\def\H{N$_{2}$H$^{+}$}
\def\D{N$_{2}$D$^{+}$}

\def\AMM{NH$_3$}

\def\WAT{H$_2$O}
\def\METH{CH$_3$OH}
\def\DMETH{CH$_2$DOH}
\def\FORM{H$_2$CO}
\def\Nfrac{$^{14}$N/$^{15}$N}
\def\cyanide{CH$_3$CN}
\def\methyl{CH$_3$CCH}

\def\formate{CH$_3$OCHO}
\def\dimethyl{CH$_3$OCH$_3$}
\def\formamide{NH$_2$CHO}
\def\ethyl{C$_2$H$_5$CN}
\def\HII{H{\sc ii}}

\def\UC{UC~H{\sc ii}}
\def\HC{HC~H{\sc ii}}
\def\UCs{UC~H{\sc ii}'s}
\def\kms{\mbox{km~s$^{-1}$}}
\def\cmc{cm$^{-3}$}

\def\solm{\mbox{M$_\odot$}}

\def\Tk{\mbox{$T_{\rm k}$}}

\def\kms{km\,s$^{-1}$}

\def\keyFont{\fontsize{8}{11}\helveticabold }
\def\firstAuthorLast{Fontani} %use et al only if is more than 1 author
\def\Authors{Francesco Fontani\,$^{1,2,3}$, Maria Teresa Beltr\'an\,$^{2}$, Anton Vasyunin\,$^{4}$}
\def\Address{$^{1}$Max-Planck-Institute for Extraterrestrial Physics (MPE), Garching bei M\"{u}nchen, Germany \\
$^{2}$Istituto Nazionale di Astrofisica (INAF), Osservatorio Astrofisico di Arcetri, Florence, Italy \\
$^{3}$Laboratory for the study of the Universe and eXtreme phenomena (LUX), Observatoire de Paris, Meudon, France \\
$^{4}$Ural Federal University, Yekaterinburg, Russia }
\def\corrAuthor{Francesco Fontani}

\def\corrEmail{francesco.fontani@inaf.it}

\begin{document}

\onecolumn
\firstpage{1}

\title[Chemical evolution in high-mass star forming regions]{Chemical evolution in high-mass star-forming regions} 

\author[\firstAuthorLast ]{\Authors} %This field will be automatically populated
\address{} %This field will be automatically populated
\correspondance{} %This field will be automatically populated

\extraAuth{}% If there are more than 1 corresponding author, comment this line and uncomment the next one.
%\extraAuth{corresponding Author2 \\ Laboratory X2, Institute X2, Department X2, Organization X2, Street X2, City X2 , State XX2 (only USA, Canada and Australia), Zip Code2, X2 Country X2, email2@uni2.edu}

\maketitle

\begin{abstract}

Growing evidence shows that 
most stars in the Milky Way, including our Sun, are born in high-mass star-forming regions, 
but due to both observational and theoretical challenges, our understanding of their 
chemical evolution is much less clear than that of their low-mass counterparts. Thanks to the 
capabilities of new generation telescopes and computers, a growing 
amount of observational and theoretical results have been recently obtained, which have 
important implications not only for our understanding of the (still mysterious) formation process
of high-mass stars, but also for the chemistry that the primordial Solar System might have
inherited from its birth environment.
In this review, we summarise the main observational and theoretical results achieved in the last decades in
the study of chemistry evolution in high-mass star-forming regions, and in the identification of chemical
evolutionary indicators. 
Emphasis is especially given to observational studies, for which most of the work has been carried out so far.
A comparison with the chemical evolution occurring in other astrophysical environments, in particular in low-mass star-forming cores and extragalactic cores, is also briefly presented.
Current open questions and future perspectives are also discussed.

\end{abstract}

\section{Introduction - the need for this review}
\label{intro}

%-----------------------------Figure Start---------------------------
\begin{figure}
% The 'scale' parameter below allows you to scale the figure so that it fits within the page. In this case the figure was scaled to 20% of its original size. Note: for .png files one has to use pdflatex, not classic latex
\includegraphics[width=13cm,angle=-90]{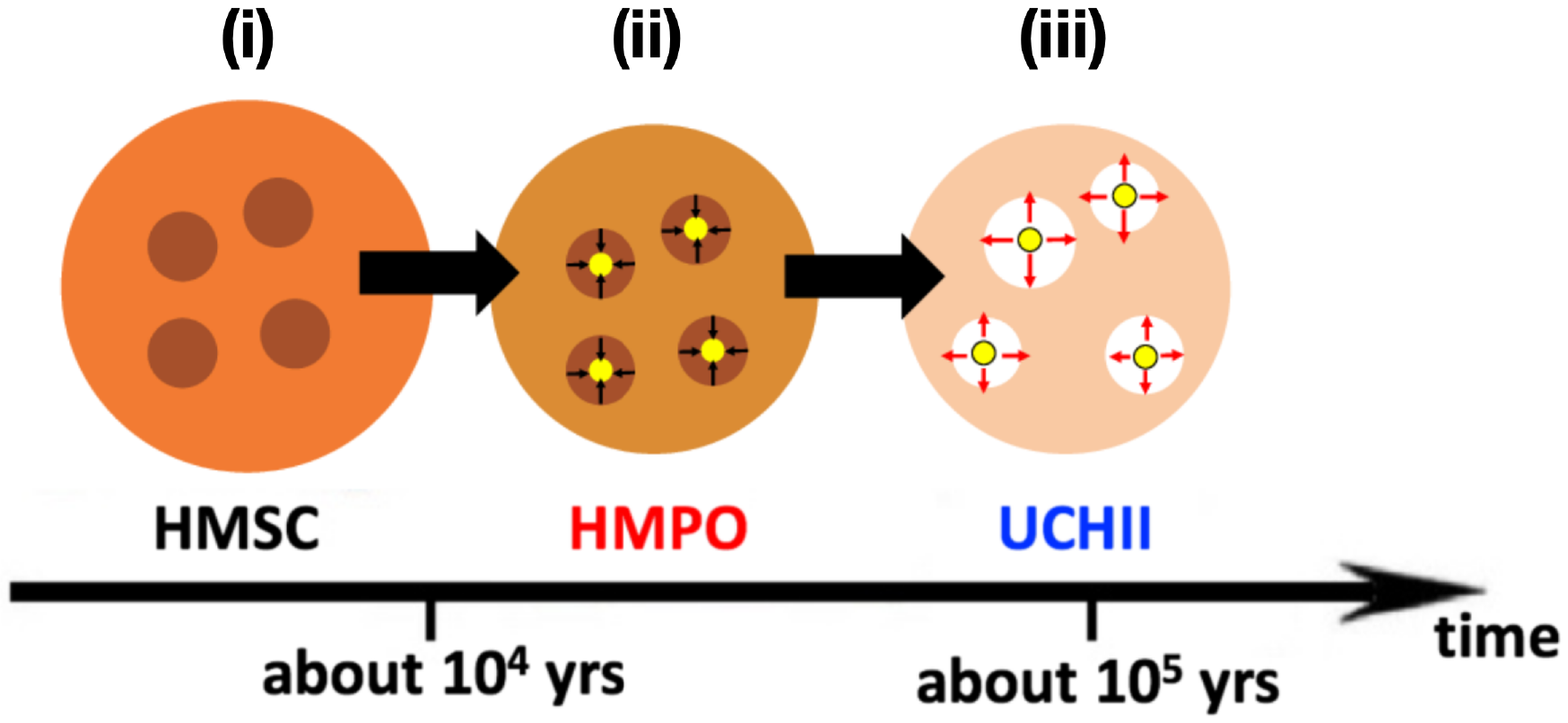}
\caption{Scheme of the coarse evolutionary classification for high-mass star-forming cores, adapted from \citet{beuther07}, following the labelling adopted in Sect.~\ref{intro}.
%The numbers above the individual stages correspond to the 6 physical evolutionary phases proposed
%by \citet{motte18}: starless (2) and protostellar (3) massive dense core; infrared-quiet (4) and infrared-bright (5) high-mass protostar(s); \HII\ region phase (6). 
Adapted from \citet{colzi20}.
}
\label{fig:Fig1}
\end{figure}
%-----------------------------Figure End------------------------------
Today, increasing evidence supports the idea that our Sun was born in a crowded stellar cluster that included stars more massive than $\sim 8$~\solm\ \citep{adams10}.  
The traces of the interaction
between these stars and the Solar System are recorded in meteoritic material, where
anomalous high abundances of daughter species of Short-Lived Radionuclides (SLRs),
in particular $^{26}$Al, produced by nearby high-mass stars during the primordial evolution 
of the Solar System, have been measured \citep[e.g.][]{portegies18,portegies19}. 
In fact, it is well-known that most stars are born in rich clusters
\citep[e.g.][]{carpenter00,lel03}
which likely included high-mass stars \citep[e.g.][]{rivilla14}.
Therefore, the study of the chemical content of high-mass star-forming regions can give 
us important information not only about the formation of high-mass stars, but also on the 
chemical heritage and complexity of both the Solar System and most stars in the Milky Way. 

Despite this, for long our knowledge of how chemistry evolves in these regions has 
remained limited. 
This is due to both observational and theoretical difficulties. 
First, observations
are challenging because high-mass star-forming cores, i.e. molecular compact structures with 
masses $\sim 100$~\solm, which have the potential to form high-mass stars and 
clusters, are fewer than low-mass cores. 
Therefore, they are on average located further away from the Sun (at typical distances larger than 1~kpc), thus smaller in angular size and typically surrounded by large amounts of ambient gas and/or nearby star formation activity difficult to disentangle \citep{motte18}. 
Second, theory predicts that massive star-forming cores evolve on timescales 
of $\leq 10^5$ years, typically shorter than those of their low-mass counterparts \citep{mckee02}.
These short timescales imply that it is challenging to define a physical evolutionary
sequence divided into well separated classes, and although tentative evolutionary 
classifications have been proposed \citep{beuther07,tan14,motte18}, their phases are debated and not clearly separated 
as in the low-mass case. 
Finding a {\it chemical evolutionary sequence} 
would be extremely useful to better understand a {\it physical evolutionary sequence}, still uncertain.
However, this is not easy because the formation and destruction of molecules 
may require timescales comparable to or longer than a single phase (i.e.~$\geq 10^4-10^5$ yrs), challenging the identification of reliable indicators of a specific evolutionary phase. 
Thanks to the improved observational and computational capabilities of new facilities, a huge and growing observational and theoretical effort has been devoted to this topic in recent years, and important results have been obtained in the last decades. 
This review article summarises the observational and theoretical endeavour devoted to 
understanding how chemistry evolves in massive star-forming regions and to identify robust chemical evolutionary indicators. 

As stated above, several attempts to empirically classify high-mass star-forming cores in different stages have been 
proposed, which can all be tentatively summarised in three coarse phases (see Fig.~\ref{fig:Fig1}): 

\begin{itemize}
\item [(i)] High-Mass Starless Cores (HMSC):
these objects, found mostly in infrared-dark, dense, molecular clouds \citep[IRDCs,][]{perault96}, 
are in a phase immediately before, or at the very beginning of, the gravitational collapse. 
They are characterised by low temperatures ($\sim 10-20$~K), masses $\geq 100$~\solm, and high densities
($n\geq 10^{4}-10^{5}$~\cmc), and do not show clear signs of on-going star formation like 
infrared sources or protostellar outflows. 
In these early cold phases, atoms and simple molecules are thought to freeze-out on dust grain surfaces, hence surface 
chemistry is very active and gas-phase chemistry is inhibited, in particular neutral-neutral
and endothermic reactions;
\item [(ii)] High-Mass Protostellar Objects (HMPOs): 
collapsing cores with evidence of one (or more) deeply embedded protostar(s), identified either by strong outflows and/or infrared objects, and characterised typically by higher densities and temperatures 
($n\simeq 10^6$~\cmc, $T\geq 20$~K). 
In this warmer environment, the molecules in the mantles sublimate back in the gas-phase, and reactions that were not efficient at low temperatures start to proceed
and form new (especially more complex) molecules. Also, collimated jets and molecular outflows from the protostar(s) can trigger locally chemistry typical of shocked gas.
\item [(iii)] Hyper- and Ultra-compact \HII\ regions: 
these objects are Zero-Age-Main-Sequence stars still embedded in the natal cloud.
Their ionising UV photons create an expanding \HII\ region. 
The densities and temperatures of the molecular 
cocoon surrounding the \HII\ region(s) ($n\geq 10^5$~\cmc, $T\sim 20 - 100$~K) can be affected 
by its (their) progressive expansion and by heating and irradiation from the central star(s).
\end{itemize} 

In our review, we will refer mostly to this coarse classification, bearing in mind that this scheme is not rigid, because these phases may partly overlap, due to the short evolutionary timescales.
Moreover, inside each group there can be a variety of physical parameters (see above), linked to possible chemical diversity.
In addition, the way in which high-mass cores can be observationally classified into the three groups can slightly change from one study to another.

\section{Observations}
\label{observations}

In this Section, we will review the main observational works that have studied
the chemical content of massive star-forming regions, to find a link between the early and late evolutionary stages. 
The approach has traditionally been twofold: line surveys towards selected objects, 
%(Belloche et al.~2013; PRIMOS, Zaleski et al.~2013;
%SPARKS, Csengeri et al.~2019; ReMoCA, Belloche et al.~2019; GUAPOS, Mininni et al.~2020), 
or selected molecular species or lines towards source surveys. 
%(e.g.~MALT90, Foster et al.~2011; TopG\"{o}t, Mininni et al. in prep.). 
We divide this section into two main subsections according to this twofold approach.
To help the reader, in Table~\ref{tab:acronyms} we list the major acronyms used for instruments, research institutes, and observational surveys mentioned throughout this section.

 \begin{table}[]
    \centering
    \begin{tabular}{llc}
    \hline
     acronym   & full definition  & type$^{(a)}$ \\
     \hline
     ALMA & Atacama Large Millimeter Array & T \\
     APEX & Atacama Pathfinder EXperiment & T \\
     ASTE & Atacama Submillimeter Telescope Experiment & T \\
     ATCA & Australia Telescope Compact Array & T \\
     BIMA & Berkeley-Illinois-Maryland Association Array & T \\
     CARMA  & Combined Array for Research in Millimeter-wave Astronomy & T \\
     CHEMICO & CHemical Evolution in MassIve star-forming COres & S \\
     CHEMOUT & CHEMical complexity in the OUTer Galaxy & S \\
     CHESS & Chemical HErschel Surveys of star-forming regions & S \\
     CoCCoA & Complex Chemistry in hot Cores with ALMA & S \\
     EMoCA  & Exploring Molecular Complexity with ALMA & S \\
      GBT & Green Bank Telescope & T \\
      GOTHAM & GBT Observations of TMC-1: Hunting Aromatic Molecules & S \\
      GUAPOS & G31.41+0.31 Unbiased ALMA sPectral Observational Survey & S \\
      HEXOS  & Herschel Observations of EXtra Ordinary Sources & S \\
      IRAM  & Institute de radioastronomie millim\'etrique & I \\
      ISO    & Infrared Space Observatory & T \\
      JCMT   & James Clerk Maxwell Telescope & T \\
      JOYS   & JWST Observations of Young protoStars & S \\
      JWST   & James Webb Space Telescope & T \\
      MALT90 & Millimetre Astronomy Legacy Team 90 GHz & S \\
      ngVLA & Next Generation Very Large Array & T \\
      NMA    & Nobeyama Millimeter Array & T \\
      NOEMA  & Northern Extended Millimeter Array & T \\
      NRAO   & National Radio Astronomy Observatory & I \\
      OSO    & Onsala Space Observatory & T \\
      PdBI   & Plateau de Bure Interferometer & T \\
      PILS   & Protostellar Interferometric Line Survey & S \\
      PRIMOS & Prebiotic Molecules Toward Sgr B2N & S  \\
      QUIJOTE & Q-band Ultrasensitive Inspection Journey to the Obscure TMC-1 Environment & S \\
      ReMoCa & Re-exploring Molecular Complexity with ALMA & S \\
      SCAMPS & SCUBA Massive Pre/Protocluster core Survey & S \\
      SEST & Swedish-ESO Submillimetre Telescope & T \\
      SMA & Sub-millimeter array & T \\
      SPARKS & Search for High-mass Protostars with ALMA up to kilo-parsec scales & S \\
      VLA  & Very Large Array & T \\
      \hline
    \end{tabular}
    \caption{Acronyms of telescopes, research institutes, and observational surveys mostly used in Sect.~\ref{observations}. $^{(a)}$ I = institute; S = survey; T = telescope.}
    \label{tab:acronyms}
\end{table}

\subsection{Line surveys of selected sources}
\label{survey-sources}

Line surveys of selected sources, and in particular unbiased spectral line surveys, have historically been the preferred approach to reveal new species in the interstellar medium (ISM). 
For example, urea detection in Sagittarius B2(N) (hereafter Sgr B2(N)) with the Re-exploring Molecular Complexity with ALMA (ReMoCA) survey \citep{belloche19} or the detection of polycyclic aromatic hydrocarbons (PAHs) in the cold, low-mass starless core TMC-1 with the GBT Observations of TMC-1: Hunting Aromatic Molecules (GOTHAM) survey \citep{mcguire21} and the Yebes 40-m telescope Q-band Ultrasensitive Inspection Journey to the Obscure TMC-1 Environment (QUIJOTE) project \citep{cernicharo21}. 
Unbiased surveys used to be very telescope time consuming, especially because the observation of the whole receiver bandwidth required several spectral setups. 
This translated in many hours of telescope time, distributed in different days or even epochs (i.e. astrophysical times used as references for some varying astronomical quantities) that led to different observing conditions and sometimes spectral noise. 
This has changed in recent years. Nowadays a new generation of new broadband receivers with much larger instantaneous bandwidth, such as those of the Submillimeter Array (SMA) or of the single-dish telescopes Institute de radioastronomie millim\'etrique (IRAM) 30-m, Yebes 40-m, or Green Bank Telescope (GBT), allow this kind of observations to be carried out faster and in a more efficient way. 
This has led to an increase of the detection rate of new species in recent years. 
According to the Cologne Database for Molecular Spectroscopy (CDMS)\footnote{https://cdms.astro.uni-koeln.de}, there are about 330 molecules detected in the ISM or circumstellar shells, and about 150 are complex organic molecules (COMs), carbon-bearing molecules with at least six atoms. 
The rate of new detections per year has steadily increased from about 4 detection/year since the end of the 1960s, beginning of the 1970s, when the first COMs were detected, to about 6 detection/year since 2005 \citep{mcguire18}. 
This rate has increased considerably in recent years, thanks to the new broadband receivers, and currently with different projects observing TMC-1, the detection rate is about 20 new species per year. 
%The future is bright thanks to the new Atacama Large Millimeter/submillimeter Array (ALMA) Wideband Sensitivity Upgrade (WSU) starting in 2025--2026, which will increase the instantaneous bandwidth of the ALMA receivers and the speed of spectral line surveys by a factor of 2 to 4, as fewer tunings will be required to cover the full ALMA bands.

In this section, we will review the results obtained from broad spectral surveys towards selected objects of known evolutionary stage. 
We will follow the three evolutionary classes described in Sect.~1: (i) high-mass starless 
core candidates (Sect.~2.1.1., e.g. \citealp{pillai11}; \citealp{wang14}); (ii) high-mass 
protostellar objects (Sect.~2.1.2.), including well-known hot molecular cores (HMCs) like Orion KL \citep{tercero10}, Sgr B2 \citep{belloche13}, G31.41+0.31 \citep{beltran09}, NGC\,6334I \citep{walsh10} or proposed hot-core precursors like G328.2551$-$0.5321  \citep{csengeri19}; and (iii) hyper- and ultra-compact (HC and UC) \HII\ regions 
(Sect.~2.1.3.) like Mon\,R2 \citep{ginard12} and W51 (\citealp{watanabe17}; \citealp{rivilla16}, \citeyear{rivilla17b}). 
We will discuss the most relevant species detected in these objects, with special attention to those with prebiotic potential.

\subsubsection{High-mass starless core candidates}

\begin{figure}
\begin{center}
\includegraphics[width=12.5cm, angle=-90]{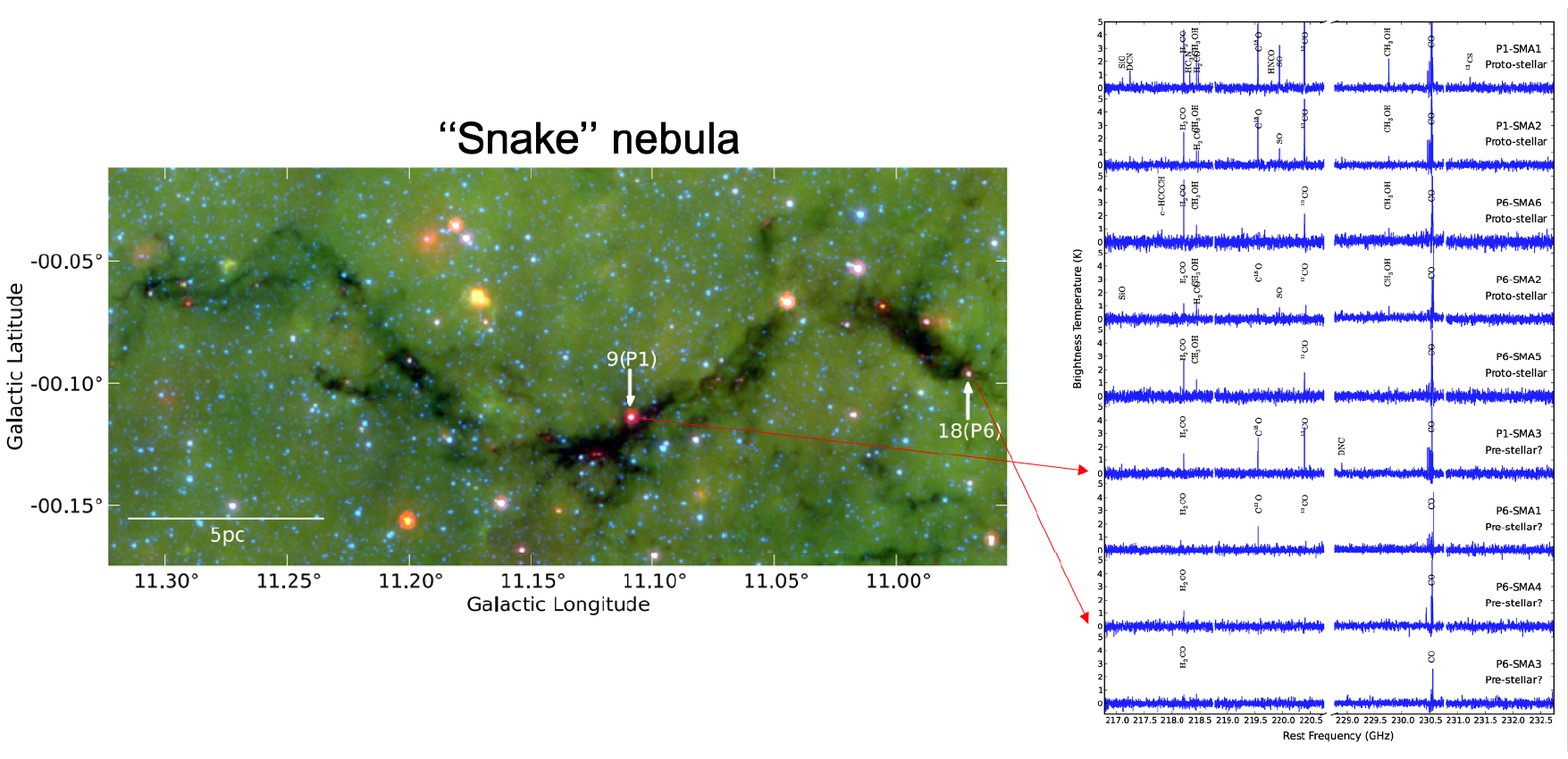}
\vspace*{-2cm}
\caption{IRDC G11.11$-$0.12. ({\it Left}) $Spitzer$ three-color  composite image (24\,$\mu$m, red, 8\,$\mu$m, green, 4.5\,$\mu$m, blue) of IRDC G11.11$-$0.12, also known as the ‘‘Snake’’ nebula. The two more massive clumps embedded in this cloud, named P1 and P6, are indicated. ({\it Right})  
SMA spectra at 1.3\,mm toward the protostellar cores and the prestellar core candidates embedded in P1 and P6, with identification of the detected species.  Adapted from \cite{wang14}.}
\label{fig:irdc-snake}
\end{center}
\end{figure}

The existence of HMSCs, typically embedded in infrared dark clouds (IRDCs), is a fundamental question in astrophysics, whose answer would allow us to discriminate and better constrain  different theories of massive star-formation, such as core accretion \citep{mckee03} and competitive accretion \citep{bonnell01}. 
The reality is that these cores are elusive and despite years of research, many of the proposed candidates have shown evidence of active star-formation, such as outflow activity (e.g., \citealp{wang11}; \citealp{tan16}; \citealp{pillai19}). 
Therefore, as a result we are left only with few starless core candidates with very cold temperatures ($<15-20$\,K) and associated with high deuterium fractionation, $R_{\rm D}$, and depletion (e.g., \citealp{pillai07}; \citealp{fontani11}; \citealp{tan13}; \citealp{wang14}, \citeyear{wang16}; \citealp{pillai11}, \citeyear{pillai19}). 
Due to the lack of a significant sample of candidates, there are not many line surveys, in particular unbiased ones, toward HMSC candidates, and most of them target specific species, in particular deuterated ones, like the SCUBA Massive Pre/Protocluster core Survey (SCAMPS) carried out with the James Clerk Maxwell Telescope (JCMT) (\citealp{thompson05}; \citealp{pillai07}) or the surveys of \cite{fontani11} and \cite{sakai12}. 
One of the few studies that has carried out a broadband spectral survey toward massive starless core candidates embedded in IRDC G11.11$-$0.12 is that of \cite{wang14}. 
The G11.11$-$0.12 cloud, also known as the ‘‘Snake’’ nebula, located at 3.6\,kpc, is one of the first IRDCs identified by \cite{egan98} and one of the best studied. 
\cite{wang14} observed the two more massive clumps embedded in this cloud, named P1 and P6 with total masses of 1.2$\times$10$^3$ and 9.3$\times$10$^2$\,$M_\odot$ and bolometric luminosities of $\gtrsim 1.3\times10^3$ and $\gtrsim1.4\times10^2\,L_\odot$, with the SMA at 1.3\,mm, covering the frequency ranges $\sim$216.8--220.8\,GHz and $\sim$228.8--232.8\,GHz, and at 850\,$\mu$m, covering the frequency ranges 333.7--337.7\,GHz and 345.6--349.6\,GHz (see Fig.~\ref{fig:irdc-snake}). 
The high-angular resolution observations have resolved the emission of the clumps into several cores, four of them being prestellar core candidates according to \cite{wang11}: cores P1-SMA3, P6-SMA1, P6-SMA3, and P6-SMA4. 
Compared with the protostellar cores embedded in P1 and P6, the prestellar cores present a much lower chemical richness,  with only very few species detected: all prestellar cores show CO and H$_2$CO and two of them also show $^{13}$CO and C$^{18}$O. 
None of them shows deuterated emission. 
\cite{wang11} have detected even fewer species toward the cores embedded in the massive P1 clump of another IRDC, the G28.34+0.06 cloud. 
In fact, these authors have detected only CO among the entire 8\,GHz SMA band toward the embedded cores, although in all cases, the CO emission is associated with molecular outflows, indicating a protostellar nature of the cores. 
Regarding IRDC G11.11$-$0.12, the poor chemistry observed in the prestellar cores contrasts with that observed in the protostellar ones.
In the latter, in addition to the emission of CO and isotopologues and H$_2$CO, there is also emission of COMs such as CH$_3$OH and $^{13}$CH$_3$OH, CH$_3$CN, CH$_3$CHO, and CH$_3$CH$_2$CN, typical of HMCs, emission of OCS, $^{13}$CS, HC$_3$N, c-HCCCH, and  HNCO, and emission of typical shock tracers such as SO and SiO. 
The richest protostellar core, P1-SMA1, also shows emission of the deuterated species DCN.  Based on the different line richness and strength of the emission, even among the protostellar cores, \cite{wang14} conclude that chemical differentiation is present in these massive clumps, likely indicating an evolutionary sequence from core to core. 

Interferometric observations with different arrays, such as the IRAM Plateau de Bure (PdBI), the Berkeley-Illinois-Maryland Association Array (BIMA), and the Very Large Array (VLA), were also carried out toward the IRDCs G29.96e and G35.20w.
Both cores are located near UC \HII\ regions in the G29.96$-$0.02 and G35.20$-$1.74 star-forming regions, respectively, as a follow-up of the SCAMPS survey by \cite{pillai11}.
The observations targeted NH$_3$, NH$_2$D, and HCO$^+$. 
The deuterated species NH$_2$D is a high-density tracer of cold  gas in very young stellar objects (YSOs). In such a cold and dense environment, carbon-bearing molecules,  particularly CO, should be depleted by freeze-out onto dust grains \citep{tafalla02}. 
This should enhance $R_{\rm D}$ due to the fact that the H$_2$D$^+$ ion, the progenitor of most deuterated species, is not destroyed by CO (see Sect.~\ref{simple}). 
Once the temperature near protostellar YSOs exceeds $\sim$20\,K, CO should be released from the surface of dust grains, react with H$_2$D$^+$, and reduce the deuterium fractionation. 
NH$_2$D was detected in 11 cores in G29.96e and 13 in G35.20w, while NH$_3$, which is a dense tracer of cold and warm gas, was detected only in 5 cores in G29.96e and 3 in G35.20w. 
The low temperatures estimated ($<20$\,K) together with the high $R_{\rm D}$, [NH$_2$D/NH$_3$] in the range 0.06--0.37, several orders of magnitude above the interstellar [D/H] ratio of $\sim10^{-5}$ \citep{oliveira03}, suggest that some of these cores could be starless.
However, the fact that HCO$^+$ molecular outflows have been observed toward both IRDCs, indicates that protostellar activity has already started in some of these cores. 
Unfortunately, the angular resolution of the HCO$^+$ observations does not allow us to distinguish which cores are indeed protostellar and which are prestellar. 
A similar study, with the same targeted species (NH$_2$D and NH$_3$) has been conducted by \cite{busquet10} in  IRAS\,20293+3952 with the PdBI and the VLA. 
This high-mass star-forming region, with a bolometric luminosity of 6300\,$L_\odot$, is associated with YSOs in very different evolutionary stages, from starless core candidates to protostellar YSOs powering molecular outflows to an \UC\ region. 
The NH$_2$D emission is strongly detected toward the prestellar candidates, with high $R_{\rm D}$ ($\sim$0.1--0.8) while it is hardly detected toward the protostellar ones, with low $R_{\rm D}$  ($<0.1$) \citep{busquet10}. 

Deuterium fractionation has also been studied in the starless core candidate C1-N and the early evolutionary phase core C1-S in IRDC G28.37+0.07 by \cite{kong16}. 
These two cores were originally identified as starless candidates based on their cold temperature ($T<20$\,K) and high $R_{\rm D}$ ([N$_2$D$^+$/N$_2$H$^+$] $\sim$0.38) by \cite{fontani11}, who carried out a N$_2$D$^+$ and N$_2$H$^+$ survey of YSOs in different evolutionary stages with the IRAM 30-m telescope. 
However, \cite{tan16} mapped in CO a narrow, highly collimated bipolar outflow associated with core C1-S, indicating that protostellar activity is already present in this core. 
Both cores have been detected in  N$_2$D$^+$, DCO$^+$, and N$_2$H$^+$,  but only C1-S in DCN (\citealp{tan13}, \citeyear{tan16}; \citealp{kong16}), and none in o-H$_2$D$^+$ \citep{kong16}.
By combining single-dish and interferometric observations, \cite{kong16} have estimated very high $R_{\rm D}$ values ($\sim$0.2--0.7) toward these two cores.

\subsubsection{High-mass protostellar objects}

Hot molecular cores, the cradles of OB stars or HMPOs, are dense, n$\gtrsim$10$^7$\,cm$^{-3}$, compact, dusty cores with
temperatures in excess of 100\,K, and luminosities $> 10^4 L_\odot$. 
These cores are often found
in association with typical signposts of massive star formation such as \UC\ regions
and maser emission of different species. 
HMCs exhibit the richest chemistry in the ISM as a result of the evaporation of the dust grain mantles by the strong radiation of the deeply embedded early-type star. 
This chemistry is mainly observable in molecular line emission at millimeter and sub-millimeter wavelengths \citep[e.g.][]{beuther07} and includes mainly hydrogenated molecules, oxygen-bearing, nitrogen-bearing, sulfur- and silicon-bearing, phosphorus-bearing, and isotopologues and deuterated species. 
Other species including chlorine, fluorine, sodium, potassium and iron have also been detected toward these massive cores. 
HMCs are the richest cores in COM emission, with species with $>12$ atoms recently detected, such as 2-methoxyethanol (CH$_3$OCH$_2$CH$_2$OH) in NGC\,6334I \citep{fried24}, and prebiotic species. 
The abundance and variety of species detected towards these regions allow us to estimate the physical parameters, such as temperature, density, and mass, and, moreover to study the kinematics of these young massive stellar objects. 

Unbiased line surveys have been carried out towards a few selected HMCs, with Sgr B2(N) in the Galactic center and Orion KL in the nearest high-mass star-forming region being the most targeted. 
The first spectral line unbiased surveys at 1\,mm and 3\,mm targeted Sgr B2 and OMC-1 (Orion\,A) in the 1980s (\citealp{johansson84} \citealp{sutton85}; \citealp{blake86}; \citealp{cummins86}; \citealp{turner89}). 
The NRAO 3\,mm survey \citep{turner89} was the only one that observed both sources and was the most sensitive one. This led to the detection of more than 700 lines in Sgr B2 and 800 in Orion\,KL \citet{turner91}. 
Some other HMCs, such as G31.41+0.31, NGC, 6334I, and NGC, 2264 CMM3, have received increasing attention, especially after the advent of ALMA. 
Next, we discuss the findings toward these sources in detail. 

{\it Sgr B2}: The Sagittarius B2 molecular cloud, located at less than 100\,pc in projected distance from the Galactic center \citep{molinari10} is the most massive star-forming region in our Galaxy. 
This giant molecular cloud complex hosts two HMCs,  Sgr B2(N) also known as the Large Molecule Heimat (LMH), and Sgr B2(M), separated by $\lesssim2$\,pc in projection and with luminosities of $\sim 10^6\,L_\odot$ and $\sim 1\times10^7\,L_\odot$, respectively \citep{schmiedeke16}, and masses of a few $\times10^4\,M_\odot$  \citep{belloche13}. 
Both HMCs contain a large number of \UC\ regions, X-ray sources and molecular masers \citep[e.g.][and references herein]{schmiedeke16} that highlight the extreme environment of this cloud. 
The Sgr B2 molecular cloud is the most chemically rich region in the Galaxy, and historically it has been the cloud in which to search for new species, being the target of numerous spectral line unbiased surveys, as already mentioned. 
Focusing on more recent surveys, the two HMCs in the region have been observed with single-dish telescopes, such as with the GBT at centimeter wavelengths, as part of the GBT Legacy Survey of Prebiotic Molecules Toward Sgr B2N (PRIMOS), the IRAM 30-m telescope at millimeter wavelengths, and the {\it Herschel} space telescope at infrared wavelengths (e.g., \citealt{neill12}, \citeyear{neill14}; \citealp{loomis13}; \citep{zaleski13};  \citealt{belloche13}; \citealt{moller21}). 
They were also observed at high-angular resolution with interferometers such as SMA \citep{qin11}, the Australia Telescope Compact Array (ATCA) \citep{corby15}, BIMA \citep{friedel04}, and ALMA \citep[e.g.][]{sanchez-monge17}, often as the target of large line surveys such as Exploring molecular complexity with ALMA \citep[EMoCA;][]{belloche16} and ReMoCA  \citep{belloche19}.
As a result of these observations, most COMs detected in the ISM, some of them heavy COMs with $\geq10$ atoms and some of them prebiotic, have been first detected in this cloud.
In particular, toward the HMC Sgr B2(N), which is chemically richer than Sgr B2(M) by a factor $\sim 3-4$ in terms of detected lines
(\citealp{belloche13}; \citealp{sanchez-monge17}), some important first detections include: formamide \citep[NH$_2$CHO,][]{rubin71}, acetic acid \citep[CH$_3$COOH,][]{mehringer97}, the first sugar-like molecule, glycolaldehyde \citep[CH$_2$(OH)CHO,][]{hollis00}, propenal (CH$_2$CHCHO) and propanal \citep[HC(O)CH$_2$CH$_3$,][]{hollis04}, the first keto ring molecule detected in a interstellar cloud, acetamide \citep[CH$_3$CONH$_2$][]{hollis06a}, cyclopropenone \citep[c-H$_2$C$_3$O,][]{hollis06b}, aminoacetonitrile \citep[NH$_2$CH$_2$CN,]{belloche08}, ethyl formate \citep[C$_2$H$_5$OCHO,][]{belloche09}, E-cyanomethanimine \citep[E-HNCHCN,][]{zaleski13}, ethanimine  \citep[CH$_3$CHNH,][]{loomis13}, which could play a role in the formation of the amino acid alanine, or urea \citep[NH$_2$C(O)NH$_2$][]{belloche19}, which like formamide contain the peptide-like bond (NCO), key for prebiotic chemistry.  
\cite{belloche16}, thanks to the EMoCA ALMA survey at 3\,mm, have reported the detection of deuterated methyl cyanide CH$_2$DCN and the tentative detection of a few other deuterated COMs (e.g., CH$_2$DOH, CH$_2$DCH$_2$CN, and CH$_3$CHDCN) toward Sgr B2(N), and have estimated a low deuteration fraction of $0.4\%$ if compared to the values estimated toward other HMCs such as Orion\, KL \citep{gerin92} and low-mass hot corinos such as 16293$-$2422 \citep{taquet14}. 
According to the authors of this study, the low deuteration fraction could be the result of the higher kinetic temperatures typical of the Galactic center and that could reduce the degree of deuterium fractionation at the end of the prestellar phase as suggested by chemical models (e.g., \citep{taquet14}). Alternatively, it might be due to an overall lower abundance of deuterium in the Galactic center.
Besides COMs, Sgr B2 is also the cloud in which some iron-bearing, chlorine-bearing, and fluorine-bearing species have been first (tentatively) detected in the ISM. 
\cite{walmsley02} reported a tentative detection of iron monoxide (FeO) seen in absorption at 1\,mm with the IRAM 30-m telescope that later on was confirmed by \cite{furuya03} with Nobeyama Millimeter Array (NMA) interferometric observations. Hydrogen fluoride (HF) was detected also in absorption with the Infrared Space Observatory (ISO) by \cite{neufeld97}, who proposed HF as the dominant reservoir of gas-phase fluorine in this source. Chlorine-bearing species such as chloronium (H$_2$Cl$^+$) has also been detected in absorption toward Sgr B2, in this case with the $Herschel$ space telescope \citep{lis10}. The star formation rate (SFR) of Sgr B2 estimated by \cite{belloche13} is 0.04\,$M_\odot$\,yr$^{-1}$. This SFR is much higher than that estimated toward other star-forming regions (see e.g., \citealp{beltran13}) and is about 2--3\% of the global SFR $1.2\pm0.2\,M_\odot$ of the Milky Way \citep{lee12}. This indicates that this cloud is a mini-starburst region. 
This, together with its location close to the Galactic center makes the physical conditions in that environment quite extreme, which could have consequences on the chemistry. Therefore, despite the wealth of new detections in this cloud, the chemistry of Sgr B2 may not be representative of that typical of HMCs in the disk of the Galaxy.

{\it Orion\, KL}: The Kleinmann-Low cloud, also known as the Orion KL cloud, which is part of the OMC-1 complex, is the closest \citep[$\sim$415 pc,][]{menten07}, most well studied high-mass star-forming region in our Galaxy, and one of the richest molecular reservoirs known. 
The cloud is very complex and contains several strong IR and millimeter sources, such as IRc2, the Becklin–Neugebauer object (BN), source\,n, and source\,I  (SrcI). The latter source, with a bolometric luminosity of $\sim10^5\,L_\odot$ \citep{greenhill04}, is, together with BN, one of the most massive objects in the region \citep[$\sim6\times10^4\,M_\odot$, ][]{ginsburg18}. 
SrcI is associated with an \UC\ region and with typical signposts of massive star formation, such as H$_2$O and SiO masers, which likely trace a massive disk (\citealp{matthews10}; \citealp{hirota12}). 
The first unbiased line surveys toward Orion KL were conducted in the 1980s with different single-dish telescopes and in various bands, from $\sim70$ to $\sim$360\,GHz,
with the Onsala Space Observatory (OSO, $\sim 72-91$~GHz) 20m telescope,  the Owens Valley Radio Observatory (OVRO, $\sim 215-263$~GHz) 10.4m telescope, and the National Radio Astronomy Observatory (NRAO, $\sim 70-115$~GHz, $\sim 200-205$~GHz, and $\sim 330-360$~GHz) 11m telescope
(\citealp{johansson84}; \citealp{sutton85}; \citealp{blake86}; \citealp{jewell89}; \citealp{turner89}).  
In more recent years, line surveys have been conducted at  (sub-)millimeter wavelengths, mainly with single-dish telescopes larger than those used in the '80s (e.g., Caltech Submillimeter Observatory, James Clerk Maxwell Telescope, GBT, IRAM 30-m, Effelsberg 100-m, Nobeyama 45-m Tianma 65-m), thus probing smaller linear scales in the frequency bands previously observed.
The frequency range covered goes from $\sim26$ to $\sim900$\,GHz(\citealp{schilke01}; \citealp{white03}; \citealp{comito05}; \citealp{goddi09}; \citealp{tercero10}; \citealp{esplugues13}; \citealp{gong15}; \citealp{rizzo17}; \citealp{suzuki18}; \citealp{liu22}, \citeyear{liu24}), providing lines with a wide range of excitation conditions. 
Orion\,KL has also been surveyed with interferometers such as CARMA (\citealp{friedel17}), SMA (\citealp{beuther05b}), and it has been selected for ALMA Science Verification observations of Band 6 (214 to 246\,GHz frequency range) and, more recently, of Band 1 (from 35 to $\sim$50\,GHz). At infrared wavelengths, Orion\,KL has been observed by the ISO space telescope from 44 to 188\,$\mu$m \citep{lerate06} and by $Herschel$ from 480 to 1907\,GHz as part of the {\it Herschel} Observations of EXtra Ordinary Sources (HEXOS) guaranteed time key program (\citealt{bergin10}; \citealt{crockett10}, \citeyear{crockett10}). 
These line surveys have detected many molecular species, especially COMs, confirming the richness of Orion\,KL, and have discovered new species in the ISM or in a star-forming region: e.g., COMs, such as methyl acetate (CH$_3$COOCH$_3$) and the {\it gauche} conformer of ethyl formate, CH$_3$CH$_2$OCOH  (\citealp{tercero10}), salty species, such as NaCl, KCl, and their isotopologues (first time detected in a star-forming region by \citealp{ginsburg19}), and new maser species, such as the vibrationally excited H$_2$O maser (\citealp{hirota12}). Several studies have indicated that the complex
chemistry of the Orion\,KL region displays chemical differentiation, where the distribution of complex nitrogen-bearing molecules is spatially different from that of oxygen-bearing molecules. 
In this scenario, nitrogen-bearing molecules would trace the HMC while oxygen-bearing molecules would trace the Orion Compact Ridge (e.g., \citealp{suzuki18}). 
However, \cite{friedel12} suggest that while for some species, such as acetone  (CH$_3$COCH$_3$), chemical processes could be responsible for the different spatial distribution, for other COMs, such as ethyl cyanide (C$_2$H$_5$CN) and methyl formate (CH$_3$OCHO), the spatial distribution will instead be determined by local physical conditions. 

\begin{figure}
% The 'scale' parameter below allows you to scale the figure so that it fits within the page. In this case the figure was scaled to 20% of its original size. Note: for .png files one has to use pdflatex, not classic latex
\begin{center}
\includegraphics[width=13cm]{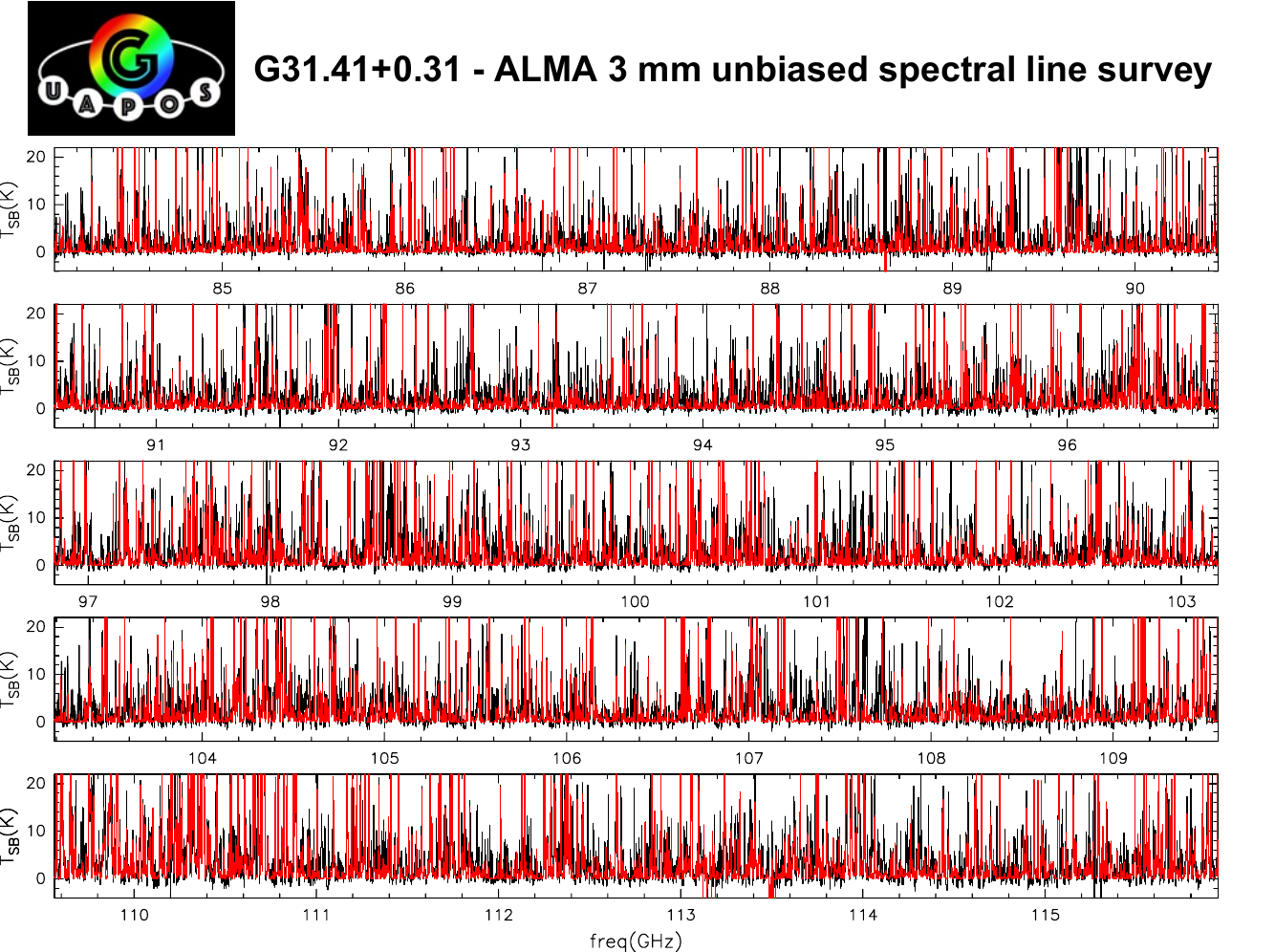}
\caption{GUAPOS survey. Full ALMA Band 3 (3\,mm) spectrum (black histogram) of the G31.41+0.31 HMC from 84\,GHz to 116\,GHz adapted
from \cite{mininni20} and \cite{colzi21}. The red solid line is the best fit for all the species detected.}
\label{fig:guapos}
\end{center}
\end{figure}

{\it G31.41+0.31}: This is a well-known HMC located at 3.75\,kpc \citep{immer19}, with a bolometric luminosity of $\sim5\times10^4\,L_\odot$ \citep{osorio09}, which does not have a \UC\ region embedded \citep{cesaroni10}. Interferometric studies at millimeter wavelengths with the IRAM Plateau de Bure Interferometer, the predecessor of the NOrthern Extended Millimeter Array (NOEMA), SMA, and ALMA have revealed that the core is very chemically rich (\citealp{beltran05}, \citeyear{beltran09}, \citeyear{beltran18}; \citealp{rivilla17a}, presenting prominent emission in a large number of COMs). In particular, the first detection of glycolaldehyde outside the Galactic center has been obtained toward G31.41+0.31 \citep{beltran09}, and heavy COMs such as both conformers of ethylene glycol (10 atoms) have also been observed (\citep{rivilla17a}; \citealp{mininni23}). The G31.41+0.31 Unbiased ALMA sPectral Observational Survey (GUAPOS) carried out an unbiased line survey of this core of the whole ALMA Band 3 (3\,mm; see Fig.\,\ref{fig:guapos}).  This survey has identified more than 40 molecules in the region (including isotopologues and lines from vibrationally-excited states), including oxygen-, nitrogen-, sulfur-, silicon-, and phosphorus-bearing species, with some of them being COMs such as formamide, acetamide, and N-methylformamide (CH$_3$NHCHO) (\citealp{mininni18}, \citeyear{mininni23}; \citealp{colzi21}; \citealp{garcia22}; \citealp{fontani24}; \citealp{lopez-gallifa24}). For some COMs, such as the isomers methyl formate (CH$_3$OCHO) and acetic acid (CH$_3$COOH), the abundances estimated toward G31.41+0.31 are higher than those detected toward Sgr B2. Regarding phosphorus-bearing species, PN has clearly been detected while PO has only been tentatively detected  \cite{fontani24}. The fact that PN has been detected along the molecular outflows in the region  and not toward the position of the HMC, together with its association with SiO and other typical shock tracers, such as SO, confirms that PN is likely a product of shock chemistry, as previously suggested by \citep{lefloch16} and \citep{rivilla20}. The G31.41+0.31 HMC is also one of the high-mass targets of the {\it James Webb Space Telescope} (JWST) Observations of Young protoStars (JOYS) Guaranteed Time Program, which aims at tracing the physical and chemical properties of about 30 young low- and high-mass protostars with the MIRI medium resolution spectrometer (MRS) instrument and its integral field unit (IFU) between 5 and 28\,$\mu$m \citep{vandishoeck25}. One of the goals of this project is to trace  COMs, such as ethanol, methyl formate or dimethyl ether, in ices and compare their abundances with those found in gas phase at (sub-)millimeter wavelengths. Initial results on the high-mass protostars IRAS\,23385+6053 and IRAS\,18089$-$1732 have identified several COMs in ices. 
For some of them, such as CH$_3$OH, the absorption feature is clearly and unambiguously identified, in other cases ice mixtures containing absorption features of acetaldehyde (CH$_3$CHO), ethanol (CH$_3$CH$_2$OH), methyl formate (CH$_3$OCHO), dimethyl ether (CH$_3$OCH$_3$) and acetone (CH$_3$COCH$_3$) (acetone) have been identified  (\citealp{rocha24}; \citealp{vandishoeck25}).

{\it NGC\,6334I}: Another HMC that has received a lot of attention from a chemical point of view is NGC\,6334I. This is one of the sources embedded in the NGC 6334 giant molecular cloud located at a distance of 1.7\,kpc \citep{neckel78}, and is the strongest one from millimeter to far-infrared wavelengths, with a bolometric luminosity of $2.6\times10^5\,L_\odot$ \citep{sandell00}. High-angular resolution dust continuum emission observations have resolved the emission of this HMC in several millimeter sources, suggesting that this HMC hosts a protocluster \citep{hunter06}. NGC\,6334I is very chemically rich, and for the density of lines, it has been compared to Sgr\,B2(N) and Orion\,KL \citep{thorwirth03}. Unbiased spectral surveys at millimeter and submillimter have been conducted with single-dish telescopes (e.g., Swedish-ESO Submillimetre Telescope (SEST), the Atacama Pathfinder EXperiment (APEX), JCMT, Mopra) in various bands, from $\sim83$ to 810\,GHz (\citealp{mccutcheon00}; \citealp{thorwirth03};  \citealp{schilke06}).  The source has also been observed with $Herschel$ from 480 to 1907\,GHz with the Heterodyne Instrument for the Far-Infrared (HIFI) as part of the Chemical HErschel Surveys of star-forming regions (CHESS) guaranteed time key program \citep{zernickel12}. High-angular resolution narrowband line observations with SMA or ATCA  or ALMA are available (e.g., \citealp{beuther05a}; \citealp{zernickel12}; \citealp{fried24}).  \citealp{mcguireetal18}) have carried out a pilot project with ALMA Band 10 by observing the source in the frequency range $\sim$874 to $\sim$881\,GHz. Like the previous HMCs analysed, NGC\,6334I is rich in emission of COMs, including prebiotic species like glycolaldehyde (e.g., \citealp{mcguireetal18}).  In a recent work, \cite{fried24} have detected with ALMA, for the first time in the ISM, one of the heaviest COMs (13 atoms), 2-methoxyethanol (CH$_3$OCH$_2$CH$_2$OH). To secure the detection of such a heavy COM, these authors have used ALMA Band 4 observations from $\sim$130 to $\sim145$\,GHz. 

There are many more HMCs that have been observed with broadband or unbiased spectral surveys. 
As an example, NGC\,2264 CMM3 at the center of the protocluter, NGC 2264 C, was observed with the Nobeyama 45-m telescope and the Atacama Submillimeter Telescope Experiment (ASTE) 10-m telescope in the 4\,mm, 3\,mm, and 0.8\,mm bands and this led to the identification of 36 molecular species and 30 isotopologues, including many emission lines of carbon-chain molecules (e.g., HC$_5$N, C$_4$H, and CCS), deuterated species (with a deuteration fractionation ranging 1\% to 4\%), and COMs \citep{watanabe15}. 
On the other hand, \cite{vanderwalt21} observed CygX-N30 also known as  W75N (B), a star-forming region containing several millimeter cores in the Cygnus\,X molecular cloud, with the SMA  interferometer. 
They observed from 329 to 361 GHz as part of the Protostellar Interferometric Line Survey of the Cygnus X region (PILS-Cygnus), and identified 29 different molecular species and their isotopologues, including many COMs and deuterated water (HDO).
The authors claim that chemical differentiation is detected in this region, with oxygen-bearing molecular species peaking toward the inner cores and nitrogen- and sulfur-bearing species peaking toward the outer ones.
\cite{vanderwalt21} interpreted this result as being due to an evolutionary effect:
The outer cores could be more evolved than the inner ones, where the gas-phase chemistry had less time to form N- and S-bearing species, and formed predominantly O-bearing ones.

Line surveys have also been conducted toward proposed HMC precursors such as G328.2551$-$0.5321 (\citealp{csengeri18}; \citealp{bouscasse22}).  
This high-mass clump, with weak or no emission in the 21--24\,$\mu$m range and a bolometric luminosity of $1.3\times10^4\,L_\odot$,  is embedded in the MSXDC G328.25$-$00.51 dark cloud, which is located at a distance $\sim$2.5\,kpc \citep{csengeri17}. 
The source was observed with ALMA from $\sim$333 to $\sim$349\,GHz as part of the Search for High-mass Protostars with ALMA up to kilo-parsec scales (SPARKS) project by \cite{csengeri18} and with an unbiased line 2\,mm, 1.2\,mm, and 0.8\,mm with APEX by \cite{bouscasse22}. 
The SPARKS survey has detected emission from 10 different COMs toward this core, such as the oxygen-bearing species ethanol, acetone, and ethylene glycol, and the nitrogen-bearing ones vinyl cyanide, ethyl cyanide, and formamide, which is a prebiotic species. 
The spatial distribution of oxygen- and nitrogen-bearing COMs in this region is different, suggesting chemical differentiation in this core: nitrogen-bearing species peak toward the protostar and oxygen-bearing COMs being associated with two accretion shock spots located in the inner envelope \citep{csengeri19}.
Based on the association with the accretion spots, these authors suggest that the COM emission in this core is produced by the sublimation of these species, but not as a result of radiative heating of dust grains, as is usually assumed for HMCs.
Deuterated water has also been observed in this core with a distribution similar to that of the oxygen-bearing COMs methyl formate and methanol. 
On the other hand, \cite{bouscasse22} have detected 39 species plus 26 isotopologues with their unbiased millimeter survey, including a  sulfur-bearing COM, CH$_3$SH, and a high abundance of the sulfur-bearing molecular ions HCS$^+$ and SO$^+$.

\subsubsection{Hyper- and Ultra-compact \HII\ regions}

OB (proto)stars deeply embedded in molecular clouds start heating and ionizing the surrounding environment as they reach the Zero Age Main Sequence (ZAMS). As the heated region grows, the ultraviolet (UV) radiation from the young embedded star will break through the core to first become a compact and dense bubble of ionised gas, known as hyper-compact \HII\ (\HC) region \citep{kurtz05}, with a size $\lesssim$0.05\,pc and a density $\gtrsim10^6$\,cm$^{-3}$, that eventually will start expanding, becoming first an ultra-compact \HII\ (\UC) region of size $\lesssim$0.1\,pc and a density $\gtrsim10^6$\,cm$^{-3}$, and later on, an extended or giant \HII\ region, with sizes of the order of 100\,pc and densities as low a a few 10s of \,cm$^{-3}$ \citep{kurtz02}. The spectra of HC and \UC\ regions are characterised by a chemistry similar to that of photo-dissociated regions (PDRs) with forbidden atomic lines of neutral oxygen and neutral and ionised carbon and nitrogen, and isotopologues, such as [OI] at 63 and 145\,$\mu$m, [CII] and [$^{13}$CII] at 158\,$\mu$m, [NII] at 205\,$\mu$m, and [CI] at 370 and 609\,$\mu$m, as well as J-ladder emission of CO and isotopologues (e.g., \citealp{huang99}; \citealp{rodon10}; \citealp{ossenkopf13}; \citealp{anderson19}; \citealp{kirsanova20}), and the presence of recombination lines (RRLs) from centimeter to ultraviolet wavelengths with very broad widths, of the order of 30--40\,km\,s$^{-1}$, or even wider \citep{kurtz05}. 

Unbiased spectral line surveys of selected HC and \UC\ have been carried out mainly at centimeter and millimeter wavelengths with single-dish telescopes (e.g., \citealp{bell93}; \citealp{ginard12}; \citealp{li17}; \citealp{watanabe17}; \citealp{liu22}) and at infrared wavelengths with $Herschel$ (e.g., \citealp{rodon10}). \cite{bell93} carried out a survey at centimeter wavelengths (between $17.6$ to $22$\,GHz) of part of the W51 \HII\ complex, including the well-known HC and \UC\ regions W51\,IRS1, W51e1, and W51e2 regions with the NRAO 43-m telescope. This survey detected 94 hydrogen (H) or helium (He) broad RRLs, as well as carbon chain molecules such as HC$_3$N and HC$_5$N, H$_2$C$_4$, C$_3$H$_2$ and its isotopologue CC$^{13}$CH$_2$ and the sulfur-bearing COM CH$_3$SH. Similar species, including CN, C$_2$H, CH$_3$CCH, different sulfur-bearing molecules such as CS and isotopologues, OCS, and SO, and recombination lines have been observed more recently by \cite{watanabe17} thanks to their spectral line survey at 3\,mm (between 85.1--101.1\,GHz and between 107.0--114.9\,GHz) carried out with the Mopra 22-m telescope. It is also worth mentioning that \cite{rivilla16} report the first detection of the prebiotic important species phosphorus monoxide, PO, toward W51e1 and W51e2.

\begin{figure}
\begin{center}
\includegraphics[width=12cm, angle=-90]{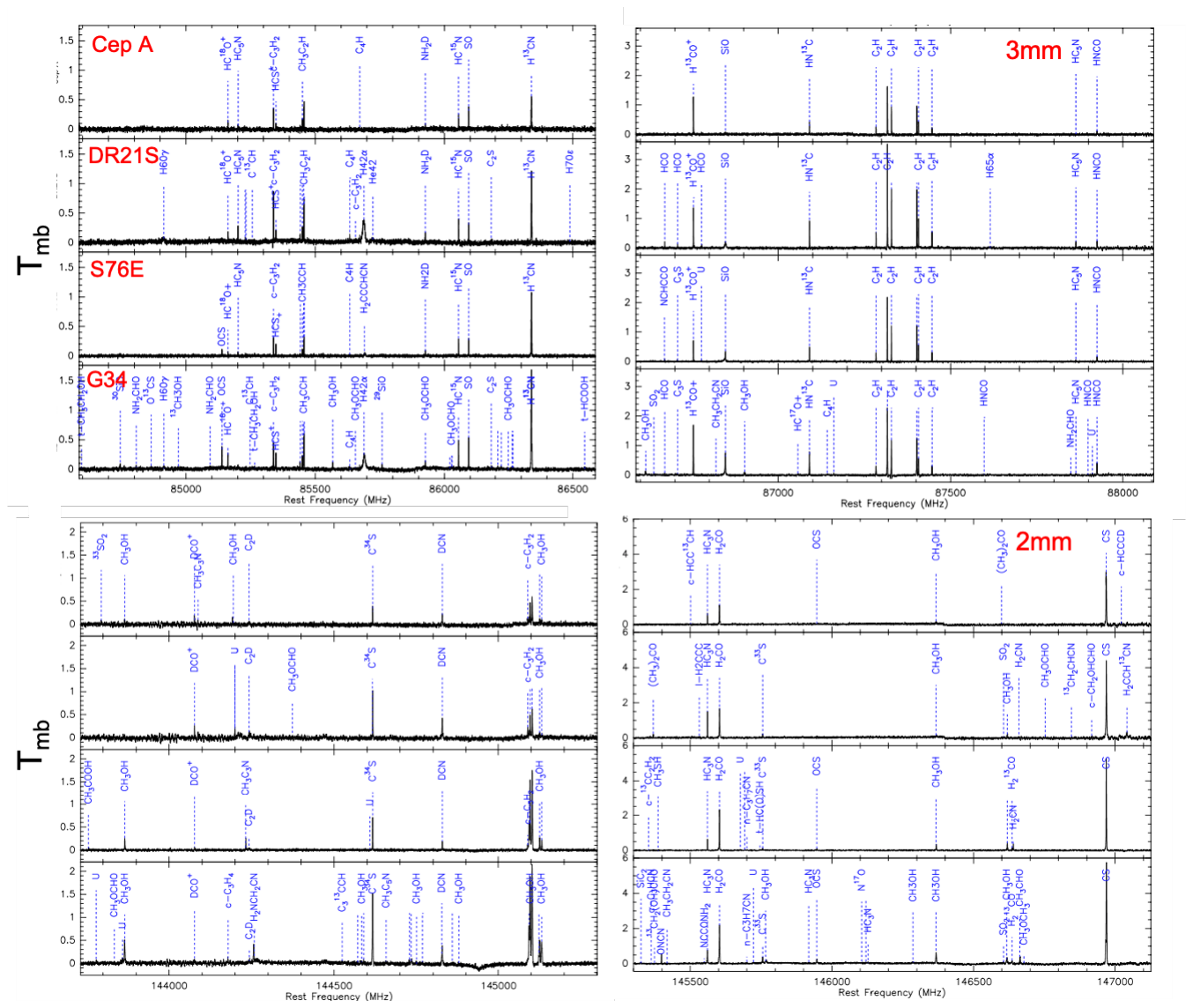}
\caption{Comparison of the IRAM 30-m spectra at 3\,mm and 2\,mm toward the \HII\ regions in Cep A, DR21S, S76E and G34. Adapted from \cite{li17}.}
\label{fig:uchii}
\end{center}
\end{figure}

One of the most observed \UC\ regions is Mon\,R2, which has been the target of spectral line surveys at 1\,mm, 2\,mm, and 3\,mm  carried out with the IRAM 30-m telescope (e.g., \citealp{ginard12}; \citealp{trevino14}). 
\cite{ginard12} detected more than 30 different species (including isotopologues) at 1\, and 3\,mm. Some of the species, such as CN, HCN, HCO, or the hydrocarbons C$_2$H, and c-C$_3$H$_2$ are typical of PDRs, although some can also be detected in HMCs (e.g, HCO and CN in G31.41+0.31;  Beltr\'an, private communication), while species such as SO$^+$ and C$_4$H confirm that UV radiation plays an important role in the chemistry of this region. 
Other species important for PDR chemistry, such as the reactive ions CO$^+$ and HOC$^+$, have been detected in Mon\,R2 and also in G29.96+0.02 by \cite{rizzo03}, who reported the first detection of these species in \UC\ regions. 
The survey of \cite{ginard12} has also revealed sulfur-bearing species, such as H$_2$CS, SO, and HC$^+$, and the deuterated species DCN and C$_2$D, with high $R_{\rm D}$ values of 0.03 and 0.05, respectively, for this kind of regions. 
On the other hand, \cite{trevino14} detected, in addition to DCN and C$_2$D, many more deuterated species (DNC, DCO$^+$, HDCO, D$_2$CO, NH$_2$D, and N$_2$D$^+$ at 1, 2, and 3\,mm, and estimated $R_{\rm D}$ values of $\sim$0.01 for HNC, HCN, C$_2$H and H$_2$CO, and $<0.001$ for HCO$^+$, N$_2$H$^+$, and NH$_3$. 
\cite{ginard12} also report the surprising detection of emission from light COMs such as CH$_3$CN, CH$_3$OH, and CH$_3$C$_2$H that is usually not detected in such environments. 
The origin of this emission is puzzling and it could be associated with dense and well-shielded cores within the molecular cloud instead of with the \UC\ region or PDR-like region. 
Obviously, RRLs have also been detected toward this region and, in particular, \cite{jimenez13} report the detection of the H30$\alpha$ at 231.9\,GHz (1.3\,mm) and H26$\alpha$ at 353.6 \,GHz (0.85\,mm) recombination line masers toward Mon\,R2, where the emission is maser amplified.   

\cite{li17} carried out a line survey with the IRAM 30m telescope from 84.5 to 92.3\,GHz at 3\,mm and from 143.7 to 147.7\, GHz at 2\,mm of four massive star-forming regions associated with \HII\ regions at different stages of evolution: Cepheus\,A (Cep A), hosting HC \HII\ regions, DR21S, associated with cometary-shaped \HII\ regions, S76E, hosting UC \HII\ regions but also with more evolved cometary-shaped and extended ring-like \HII\ regions, and G34.26+0.15 (G34), containing an extended \HII\ region (see Fig.~\ref{fig:uchii}). 
This survey has detected many different species towards these regions, including long carbon chain molecules such as HC$_{2n-1}$N and istopologues (with $n=2, 3$), CH$_3$C$_2$H, c-C$_3$H$_2$, and C$_4$H, deuterated molecules such as DCN, HCO$^+$, and NH$_2$D, and shock tracers such as SiO, SO, and HNCO toward all sources. 
Species associated with PDRs and hydrogen RRLs (H42$\alpha$ only in DR21S), were detected toward all except Cepheus\,A. 
In fact, the study concluded that the chemical richness of the sources increased with their evolution, as indicated by the fact that while 78 different molecular species plus 3 RRLs were detected in G34.26+0.15, ``only" 35 different species were identified in Cepheus\,A in the molecular envelope surrounding the HC \HII\ region.
 
In their single-dish Q-band line survey toward Orion KL, covering the 34.8--50\,GHz frequency range, \cite{liu22} also observed the well-known \HII\ region M42. A total of 126 hydrogen RRLs and 40 He RRLs, with a maximum $\Delta n$ of 16 and 7, respectively (H135$\pi$ and He99$\eta$), were detected toward this \HII\ region. The survey has also tentatively detected 11 carbon RRLs with a maximum $\Delta n$ of 3 (C81$\gamma$), but as \cite{liu22} discuss, these lines should originate from the PDRs located between M42 and the Orion Bar. 

%and the Stratospheric Observatory for Infrared Astronomy (SOFIA) airborne observatory (e.g., \citealp{kirsanova20}). 

At infrared wavelengths, \cite{rodon10} observed the optically visible  UC \HII\ region Sh2-104 with the PACS and SPIRE instruments of $Herschel$ at 100, 160, 250, 350, and 500\,$\mu$m. They detected the $J$-ladders of $^{12}$CO and $^{13}$CO up to the $J$ = 13--12 and $J$ = 9--8 transitions, respectively,  as well as proxies for ionizing flux such as ionised nitrogen [NII] and neutral carbon [C{\sc I}].  Some other authors, like \cite{kirsanova20}, have observed PDRs around \HII\ regions, and have revealed forbidden atomic lines of ionised carbon and isotopologue ([CII], [$^{13}$CII])  and neutral oxygen ([OI]), as expected for the chemistry of such regions.

\subsubsection{Summary, open questions, and future prospects}

We have analysed the main results of line surveys of selected sources in different evolutionary stages. Starting with HMSCs, as discussed above, the main problem that one encounters is the real existence of such starless cores. Many studies have revealed that previously thought starless cores embedded in IRDCs are indeed protostellar objects as indicated by the fact that they are powering molecular outflows  \citep[e.g.][]{wang11, tan13}. However, despite the fact that these cores  show already protostellar activity, their evolutionary stage is earlier than that of the well-known HMPOs discussed in Sect.~2.1.2) and their chemistry is also different. The most promising prestellar core candidates exhibit a poorly rich and very simple chemistry, with emission of simple species such as CO and isotopologues, H$_2$CO, NH$_3$ and N$_2$H$^+$ and emission of deuterated species such as DCN, NH$_2$D, and N$_2$D$^+$. In fact, the main characteristic of HMSCs is the high $R_{\rm D}$, which can be orders of magnitude above the interstellar one. 

The chemistry of slightly more evolved YSOs, i.e., protostellar cores embedded in IRDCs, is  also characterised  by the presence of deuterated species. However, the chemical richness of these protostellar cores is greater, and in fact, there spectra show that the emission of COMs, such as CH$_3$CN, CH$_3$CHO, or CH$_3$CH$_2$CN, and of shock tracers, such as SiO and SO, associated with their driven molecular outflows, is important.

The richest chemistry in the ISM, as revealed by unbiased spectral line surveys, is that of HMPOs associated with HMCs. Such chemistry is dominated by the emission of COMs, which can be very heavy or large, i.e., they can consist of many atoms (e.g., 2-methoxyethanol, CH$_3$OCH$_2$CH$_2$OH, \citealt{fried24}), and can be prebiotic, such as formamide or glycolaldehyde. The most common species observed in HMCs are oxygen-bearing, nitrogen-bearing, sulfur-bearing, and silicon-bearing species, although species containing phosphorus, iron, chlorine, fluorine, and potassium, and salty species, such as NaCl and KCl and their isotopologues, have also been detected. In some HMCs,  the spatial distribution of oxygen- and nitrogen-bearing species, particularly COMs, is different and this has been explained in some cases as the result of chemical differentiation in the core. However, it cannot be ruled out that local chemical processes or physical conditions are responsible for the different spatial distribution. The chemistry of HMPOs is also rich in emission from vibrationally excited states of (e.g., HC$_3$N $v_{1...7}=1$, CH$_3$CN $v_8=1$, and CH$_3$OH $v_t=1$), and maser emission, mainly of H$_2$O, CH$_3$OH, and OH. On the other hand, despite deuterated species of e.g., CH$_3$CN, CH$_3$OH, and H$_2$O,  have been detected toward HMCs, the deuteration fractions are low.  

The chemistry of the most evolved massive YSOs, i.e., HC and \UC\ regions, is that of highly UV-irradiated environments, and therefore, it resembles that of PDRs. Their IR spectra is rich in forbidden atomic lines of neutral oxygen and neutral and ionised carbon
and nitrogen, J-ladder emission of CO, and isotopologues. Another characteristic feature of their spectra from radio to ultraviolet wavelengths is the presence of very broad recombination lines (RRLs). 

The future of unbiased spectral line surveys is bright thanks to the new Atacama Large Millimeter/submillimeter Array (ALMA) Wideband Sensitivity Upgrade (WSU). This upgrade, which will start in 2025--26 with the new Band 2 receiver operating from 67 to 116\,GHz, will increase the instantaneous bandwidth of the ALMA receivers and the speed of spectral line surveys by a factor of 2 to 4, as fewer tunings will be required to cover the full ALMA bands. The WSU will also increase the spectral line sensitivity by improving the receiver temperatures. In summary, the WSU will advance astrochemistry studies by enabling the detection of fainter species and rapidly increasing the number of observed sources. This will improve statistics and will allow to more accurately determine the chemical complexity of massive YSOs at all evolutionary stages. 

\subsection{Source surveys of selected lines}
\label{survey-lines}

Several molecular abundances, or abundance ratios, were proposed as chemical clocks by investigating their variation in source surveys of cores divided in different evolutionary stages.
In this section, we review these studies focussing on the following type of molecules: simple molecules, less abundant isotopologues and isotopic ratios, COMs, and shock tracers.
We conclude this section with a quick overview of the studies that proposed maser emission lines as indicators of evolution.

\subsubsection{Simple molecules}
\label{simple}

Several chemical evolution studies took  \H\ as a reference species.
\H\ is one of the most abundant molecular ions, and it is clearly detected in all stages of the high-mass star-formation process with a moderate variation from the HMSC to the \UC\ phase \citep{hoq13,gerner15}.
Moreover, its progenitor species, that is N$_2$, appears to be less depleted than CO, the progenitor of another very abundant ion, HCO$^+$.
\citet{wang23} found that the \CIII/\H\ ratio increases with evolution when comparing IRAM 30m observations of 61 \UC\ regions with those 
towards HMSCs and HMPOs
previously obtained by  \citet{taniguchi19}.
The increase, mostly due to the increase of the \CIII\ abundance, was attributed to the enhanced amount in the gas of C$_2$H$_2$, the main progenitor of \CIII\ through the neutral-neutral reaction \citep{takano98}: 
\begin{equation}
   {\rm C_2H_2+CN\rightarrow HC_3N+H\;.}
\end{equation}
In contrast, \H, produced in the gas by the ion-neutral reaction ${\rm N_2+H_3^+}$, is also formed efficiently in the early cold phases.
\citet{taniguchi18} have also measured an average increase of the \CIII\ column density with evolution through observations of 17 HMSCs and 35 HMPOs.
However, its fractional abundance with respect to H$_2$ is found to decrease with time, and the \CIII\ column density decreases with the luminosity-to-mass ($L/M$) ratio of the targets, a well known evolutionary indicator \citep[e.g.][]{lopez11}.
This suggests a destruction of \CIII\ with protostellar activity.
Therefore, the use of the \CIII/\H\ ratio as evolutionary indicator is not yet fully clear.

Other proposed chemical clocks involving \H\ are \H/HCO$^+$, and \H/CCS.
The \H/HCO$^+$ ratio was investigated by \citet{yew15} using data of the Millimetre Astronomy Legacy Team 90 GHz (MALT90) survey.
MALT90 used the Mopra 22 m single-dish telescope of the Australia Telescope National Facility to map 2000 dense molecular clumps hosting different stages of the high-mass star formation, from HMSCs to \UCs, in 16 lines of dense gas tracers with frequencies in between $\sim 86.7$ and $\sim 93.2$~GHz, among which the $J=1-0$ transitions of HCO$^+$, H$^{13}$CO$^+$, HCN, HNC, HN$^{13}$C, and \H\ \citep{jackson13}.
To derive the \H/HCO$^+$ ratio, \citet{yew15} used the optically thin isotopologue H$^{13}$CO$^+$.
They report a marginal decrease of this ratio from the HMPO to the \UC\ stage, possibly due to depletion of \H\ due to the formation of the ionised expanding region.
\citet{chen25} proposed that the \H/CCS ratio increases with evolution, perhaps owing to the simultaneous increase of \H\ production and decrease of CCS abundance.
CCS is indeed a species thought to be formed in the gas-phase by atomic carbon of sulphurated ions at early stages, although \citet{fontani23} did not find a significant decrease of its abundance with the evolution in high-mass star-forming cores.
However, the studies of \citet{chen25} and \citet{fontani23} probe sources with different average heliocentric distances, and hence potentially different linear scales.

\citet{fontani21} suggested that the chemistry of HCNH$^+$, a molecular ion rare but believed to be the main gas-phase precursor of HCN and HNC (upon dissociative recombination), is different in HMSCs and more evolved cores.
In particular, the abundance ratios HCNH$^+$/HCN and HCNH$^+$/HCO$^+$ are both below 0.01 in HMPOs and \UCs, and higher than this threshold in HMSCs.
Chemical modelling suggests that this difference is due to the different formation of HCNH$^+$: in HMSCs, the main progenitors should be HCN$^+$ and HNC$^+$, while in the later stages, where the abundance of HCN$^+$ and HNC$^+$ drops by three orders of magnitude, should be  HCO$^+$ and HCN/HNC \citep{fontani21}.
Therefore, these ratios can be useful tools to identify HMSC candidates.

\subsubsection{Isotopic fractionation}
\label{fractionation}

%-----------------------------Figure Start---------------------------
\begin{figure}
\includegraphics[width=17cm]{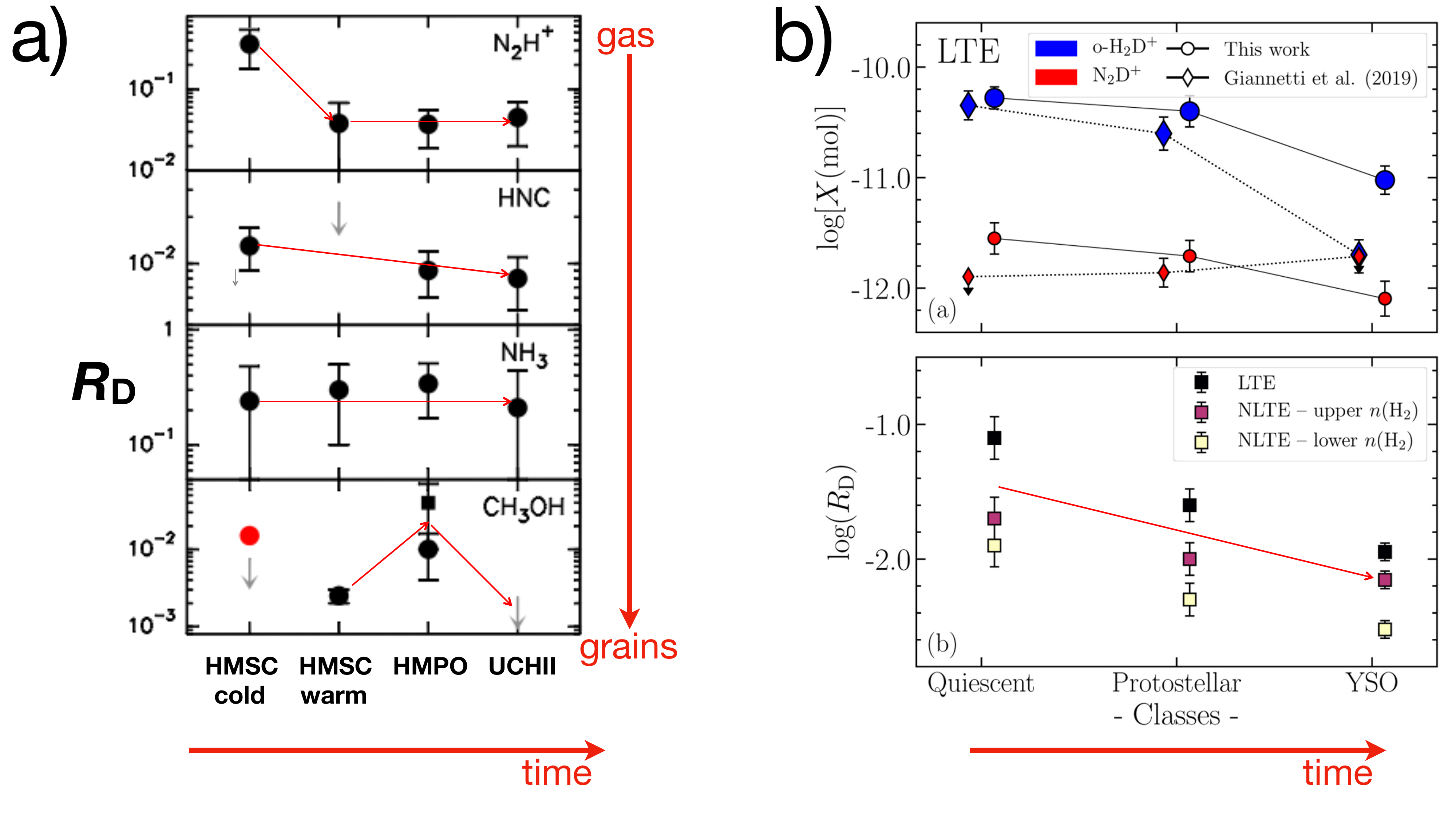}
\caption{Panel (a), from top to bottom: comparison 
between the mean D/H ratio (black symbols) computed from \H\ (first panel), HNC (second panel), \AMM\ (third panels), and \METH\ (fourth panels). 
The mean values have been computed for the evolutionary groups
HMSC, HMPO, and \UC\ (Fig.~\ref{fig:Fig1}).
Cold and warm HMSCs, namely cores with kinetic temperatures lower and higher than 20~K, respectively, have been treated separately. 
The error bars indicate the standard deviation. 
The grey arrows represent mean upper limits for those evolutionary groups
in which no sources have been detected. 
The red dot in the fourth
panel represents a doubtful \DMETH\ detection.
%Bottom panel: mean rotation temperatures (filled diamonds) derived from ammonia in the four groups. 
The red arrows in each frame illustrate roughly the tentative evolutionary trends.
Adapted from Fontani et al.~(2015a).\\
Panel (b), top: evolutionary trends observed by \citet{sabatini24} of {\it o-}H$_2$D$^+$ (blue symbols) and \D\ (red symbols) as a function of the evolutionary class. 
Circles and diamonds refer to results from \citet{sabatini24} and \citet{giannetti19}, respectively. 
Panel (b), bottom: median $R_{\rm D}$ factors derived for each evolutionary class. 
Different colours refer to values obtained from LTE and Non-LTE analysis discussed in
\citet{sabatini24}.}
\label{fig:Dfrac}
\end{figure}
%-----------------------------Figure End------------------------------

The process indicated under the name of isotopic fractionation refers to the chemical reactions that affect the distribution of the different isotopes of an element into molecules.
Under specific physical conditions, the isotopes can be unequally distributed with respect to their elemental ratio.
For example, deuterated molecules in cold and dense cores can have abundances higher than the D/H elemental ratio \citep[$\sim 2.5\times 10^{-5}$][]{zavarygin18} even by several orders of magnitude, owing to the exothermic forward reaction \citep{watson74}:
\begin{equation}
{\rm H_3^+ + HD \rightleftharpoons H_2D^+ + H_2 + 230 K} \;.  
\label{eq:fractionation}
\end{equation}
If the gas kinetic temperature, \Tk, is lower than $\sim 20$~K and H$_2$ is mainly in the para- form\footnote{The ground state of ortho-H$_2$ is $\sim 170$~K higher than that of the para-H$_2$ \citep{beyer19}. This excess energy of the ortho- form can help to overcome the exothermicity of the backward reaction \ref{eq:fractionation}.}, this reaction can proceed only from left to right, producing high H$_2$D$^+$/H$_3^+$ ratios.
Moreover, at densities higher than $\sim 10^4$~\cmc, CO, the main destruction partner of H$_2$D$^+$, is heavily depleted \citep[namely the gaseous CO molecules get frozen onto dust grains, and hence the CO gaseous molecular abundance decreases, see e.g.][]{bet07}.
CO depletion boosts reactions of H$_2$D$^+$ with other non-depleted neutrals such as, for example, N$_2$.
This explains why the {\it deuterium fraction}, $R_{\rm D}$, namely the column density ratio between a deuterated species and its hydrogenated counterpart, of non depleted molecules is claimed to be a powerful chemical diagnostic to distinguish between cold/early and warm/late evolutionary stages in star-forming cores \citep[e.g.][]{cec12,ceccarelli14}.

Observations of high-mass star-forming cores in different evolutionary stages have indeed highlighted that the \D/\H\ and the H$_2$D$^+$/H$_3^+$ ratios decrease from the early phases to the more evolved stages \citep[e.g.][]{fontani11,gerner15,giannetti19,li22,pazukhin23,sabatini24}.
Figure~\ref{fig:Dfrac} shows the observed \D/\H\ ratios measured by \citet{fontani11} (panel (a)), and by \citet{giannetti19} and \citet{sabatini24} (both in panel (b)).
In each work, the classification of the cores into different evolutionary stages is based on different criteria. 
For example, \citet{fontani11} separate the HMSC and HMPO classes based on the presence/absence of outflows and any embedded mid-infrared source, while \citet{sabatini24} consider as protostellar objects sources detected at 70~$\mu$m but still undetected at 24~$\mu$m.
\citet{fontani11} make a further distinction in the HMSC group between cold and warm cores based on a \Tk\ lower or higher than 20~K, respectively (Fig.~\ref{fig:Dfrac}).
Nevertheless, the \D/\H\ fraction shows a decreasing trend with evolution in all works, regardless of the criteria used to classify the sources, and to the analysis performed to derive the molecular column densities (e.g. using a local thermodynamic equilibrium
(LTE) or non-LTE approach).
Interestingly, the very high \D/\H\ ratio in HMSC candidates is inferred from both single-dish observations \citep[e.g.][]{fontani11,gerner15,sabatini24}, and interferometer images \citep[e.g.][]{tan13,kong16,li22}, suggesting that this behaviour does not depend on the probed angular scale.

Figure~\ref{fig:Dfrac} also compares the \D/\H\ ratio with {\it o-}H$_2$D$^+$/H$_3^+$, DNC/HNC, NH$_2$D/\AMM, and \DMETH/\METH.
While the evolutionary decreasing trend seen in the {\it o-}H$_2$D$^+$/H$_3^+$ ratio clearly resembles that obtained in \D/\H, the DNC/HNC ratio shows a marginal decrease with evolution, NH$_2$D/\AMM\ shows no significant decrease, and \DMETH/\METH\ seems to have the largest value at the HMPO stage.
The natural explanation for the similar decreasing $R_{\rm D}$ of the two molecular ions is the gas temperature enhancement with evolution, which causes the progressive destruction of H$_2$D$^+$ and of its daughter species \D\ via the backward reaction~\ref{eq:fractionation}.
The increase in the ionisation fraction during protostellar evolution can also contribute to regulate the abundance of these ions \citep[e.g.][]{socci24}.
The other species in panel (a) of Fig.~\ref{fig:Dfrac} are believed to form predominantly in the gas (DNC and HNC), or partially (NH$_2$D and \AMM) or totally (\DMETH\ and \METH) on grain ice mantles.
Ammonia, methanol, and their deuterated isotopologues, are produced efficiently on grain mantles during the early dense and cold phases through hydrogenation of N and CO, respectively. 
Then, when the nascent protostellar object heats up the surrounding gas, the evaporation of the grain mantles progressively releases these molecules into the gas.
Finally, as the HMPO evolves towards a \UC\ region, the deuterated species are expected to be gradually destroyed because of the higher efficiency of the backward endothermic gas-phase reactions.
However, while \AMM\ can also have an efficient formation route in cold gas
\citep[e.g.][]{herbst21}, \METH\ is expected to be a pure surface chemistry product.
The trend shown in Fig.~\ref{fig:Dfrac} for the \DMETH/\METH\ ratio is overall consistent with this scenario, while that of the NH$_2$D/\AMM\ ratio suggests that the different formation and destruction mechanisms balance each other out along evolution.
As for the behaviour of the DNC/HNC ratio, although DNC is believed to be a gas-phase product like \D, the destruction of DNC into the warm gas appears to be slower than that occurring to \D\ \citep{sakai12,fontani14}, and this could explain the flatter decrease of its $R_{\rm D}$ with time.

Are other isotopic ratios (e.g.~$^{14}$N/$^{15}$N, $^{16}$O/$^{18}$O, and $^{12}$C/$^{13}$C) also evolutionary indicators?
In principle, all backward fractionation reactions as that in Eq.~\ref{eq:fractionation} are endothermic and have temperature thresholds due to the lower zero-point energy of the molecules containing the heavier, less abundant isotope \citep{watson74}.
The zero-point energy depends on the reduced molecular mass, therefore the difference between reactants and products is more pronounced for itotopologues that have remarkable mass differences, like those containing D and H (Eq.~\ref{eq:fractionation}).
In contrast, the exothermicities of the reactions that drive the isotopic fractionation of oxygen, carbon, and nitrogen, are of an order of magnitude lower than that in Eq.~\ref{eq:fractionation}, owing to the smaller mass difference between reactants and products \citep[e.g.][]{mer14,mer17}.
Therefore, the greatest effect is expected in the D/H ratio, and the observational studies mentioned above confirmed this.
Nevertheless, several studies have investigated whether some of the other isotopic fractions are evolutionary indicators.
In this respect, the \Nfrac\ ratio is particularly intriguing because it shows a variation of an order of magnitude ($\sim 100-1000$) in low- and high-mass starless and protostellar cores, which even depends on the molecule used \citep[e.g.][]{cec12,bizzocchi13,fontani15b,daniel16,colzi18a,desimone18,redaelli20}.
\citet{fontani15b} and \citet{colzi18a} investigated the \Nfrac\ ratio in \H\ and HCN, respectively, in the evolutionary sample divided in HMSCs, HMPOs, and \UC s previously studied in deuterated molecules by \citet{fontani11} and \citet{fontani15}. 
Both studies did not show a clear trend of \Nfrac\ with evolution.
Moreover, the \Nfrac\ ratios measured from either \H\ and HNC both show a limited dispersion (\Nfrac$\sim 100 -1100$ and $\sim 250-600$, respectively) around the protosolar nebula value \citep[$\sim 441$][]{marty11}.
This indicates that the chemical evolution does not seem to play a role in the fractionation of nitrogen.
Rather, the \Nfrac\ ratio seems more sensitive to nucleosynthesis processes, that can change the isotopic ratios across the Galaxy because of the varying stellar yields \citep[e.g.][]{colzi18b,colzi22,romano22}.

\subsubsection{Carbon chains and Complex Organic Molecules}
\label{complex}

Carbon chains and COMs are essential species in astrochemistry.
Both are important interstellar reservoirs of carbon, and precursors of prebiotic molecules \citep[e.g.][]{cec12,jorgensen20}, but while carbon chains are thought to be formed from atomic (neutral or ionic) carbon, abundant in the early stages of star formation when C is not yet locked almost totally in CO, the emission of COMs characterises especially the evolved stages.
Observations of some carbon chain species (C$_2$H, CCS, {\it c-}C$_3$H$_2$, \CIII, \CV) were carried out by \citet{taniguchi18} and \citet{taniguchi19} using the Nobeyama 45m telescope towards a sample of HMSCs and HMPOs.
As mentioned in Sect.~\ref{simple}, they proposed the \H/\CIII\ as evolutionary tracer because associated with a decreasing trend with evolution, due to simultaneous \H\ destruction upon reaction with CO, and \CIII\ production through warm carbon chain chemistry \citep{sey13}.
\citet{fontani25} studied the evolution of several carbon-bearing species through the observational project 'CHemical Evolution in MassIve star-forming COres (CHEMICO)'.
The project, through an unbiased spectral survey obtained with the IRAM 30m telescope, aims at investigating
any aspect of the chemical evolution of high-mass star-forming cores by observing representatives of the three main evolutionary
categories.
They found that carbon chains and hydrocarbons tend to trace gas at lower temperatures than cyanide species such as \CIII, \cyanide, and \CV.
Moreover, the only species whose measured fractional abundances show a significant enhancement with evolution are the COMs \methyl\ and \cyanide.

COMs are believed to be formed either through gas-phase chemical reactions \citep[e.g.][]{dew85,veh13,balucani15,skouteris18}, and surface chemistry processes \citep[e.g.][]{hasegawa92,reh00,garrod08,ruaud15}. 
These processes are not completely independent and can interplay \citep[e.g.][]{hasegawa92,balucani15}, as we will discuss in Sect.~\ref{theory}.
However, no matter what the formation pathway is, their abundance in the gas-phase is expected to increase with temperature, and thus evolution, because higher temperatures trigger endothermic gas-phase reactions and cause grain mantle evaporation.
The observational studies conducted so far agree with this general picture, although COMs have also been detected in IRDC cores \citep[e.g.][]{vasyunina14,beaklini20}.
\citet{gerner14} studied the chemical content of a sample of 59 high-mass star-forming cores divided in the evolutionary groups in Fig.~\ref{fig:Fig1}, but with the further distinction, in the HMPO group, between early chemically poorer objects and later HMCs characterised by richer chemistry.
The work focussed on simple molecules, but included also the COMs \METH\ and \cyanide, and found for both a progressive increase in the molecular abundance from the HMSC to the HMC stage.
\citet{coletta20} surveyed 39 high-mass star-forming cores.
This sample includes the 27 sources selected and studied by \citet{fontani11}, and additional 9 objects classified as \UC s or intermediate between the HMPO and \UC\ stage.
The species studied were \formate, \dimethyl, \ethyl, and \formamide.
The fractional abundances of all these species clearly increase with evolution, covering 6 orders of magnitude in the $L/M$ ratio.
Using ALMA, the project 'Complex Chemistry in hot Cores with ALMA' \citep[CoCCoA,][]{chen23,chen25b} surveyed 14 high-mass chemically rich star-forming regions in several oxygen-bearing COMs.
Among these, acetone (CH$_3$COCH$_3$) is found to be a peculiar species, because the abundance ratio acetone-to-methanol is lower in the gas phase than in ices by an order of magnitude \citep{chen24}.
This suggests gas-phase destruction after sublimation from grain mantles, and indicates that care needs to be taken when using COMs as evolutionary indicators, because important gas-phase destruction pathways in the late evolutionary stages could be overlooked.

\subsubsection{Shock tracers}
\label{shock}

Protostellar outflows are ubiquitous in the formation process of stars of all masses \citep{lee20}, which can locally attain velocities greater than 100~\kms. 
The shock due to the passage of the outflow can destroy both the ice mantles (grain sputtering) and the refractory cores (grain shattering) of dust grains. Molecules efficiently formed on ice mantles via surface chemistry, or synthesised in the warm gas from material released from the refractory cores, are indeed greatly enhanced in abundance. 
As stated in Sect.~\ref{survey-sources}, typical selective shock tracers are Silicon- (SiO, SiS), sulphur- (SO, SO$_2$, H$_2$S), and phosphorus-bearing (PO, PN) species.
But also species that are efficiently produced on ice mantles (H$_2$O, OH, \METH , \FORM , \AMM , HCN), increase their abundance by orders of magnitude in shocks  \citep[e.g.][]{bep97,jorgensen04,lee20,codella20}.

SiO is certainly believed to be the most selective indicator of protostellar shocks.
\citet{lopez11} observed with the IRAM 30m telescope the $J=2-1$ emission in 32 infrared dark cloud clumps, and detected emission at high velocities in most of them ($88\%$).
Moreover the line luminosity is found to drop with the $L/M$ ratio, which suggests a decline of protostellar jet activity with evolution.
A different result was achieved by \citet{rodriguez23}, who observed with the VLA the SiO $J=1-0$ line towards ten jet candidates associated with high-mass protostars, and found no correlation between the line luminosity and the $L/M$ ratio.
Similarly, \citet{liu21} observed with ALMA the SiO $J=5-4$ line in 32 infrared dark cloud clumps, and found no (anti-)correlation between line intensity and the $L/M$ ratio.
These studies mention the possibility that the lack of an anti-correlation between line intensity, or line luminosity, and the $L/M$ ratio is due to a very similar age of the sources studied.

\begin{figure}
\includegraphics[width=16cm]{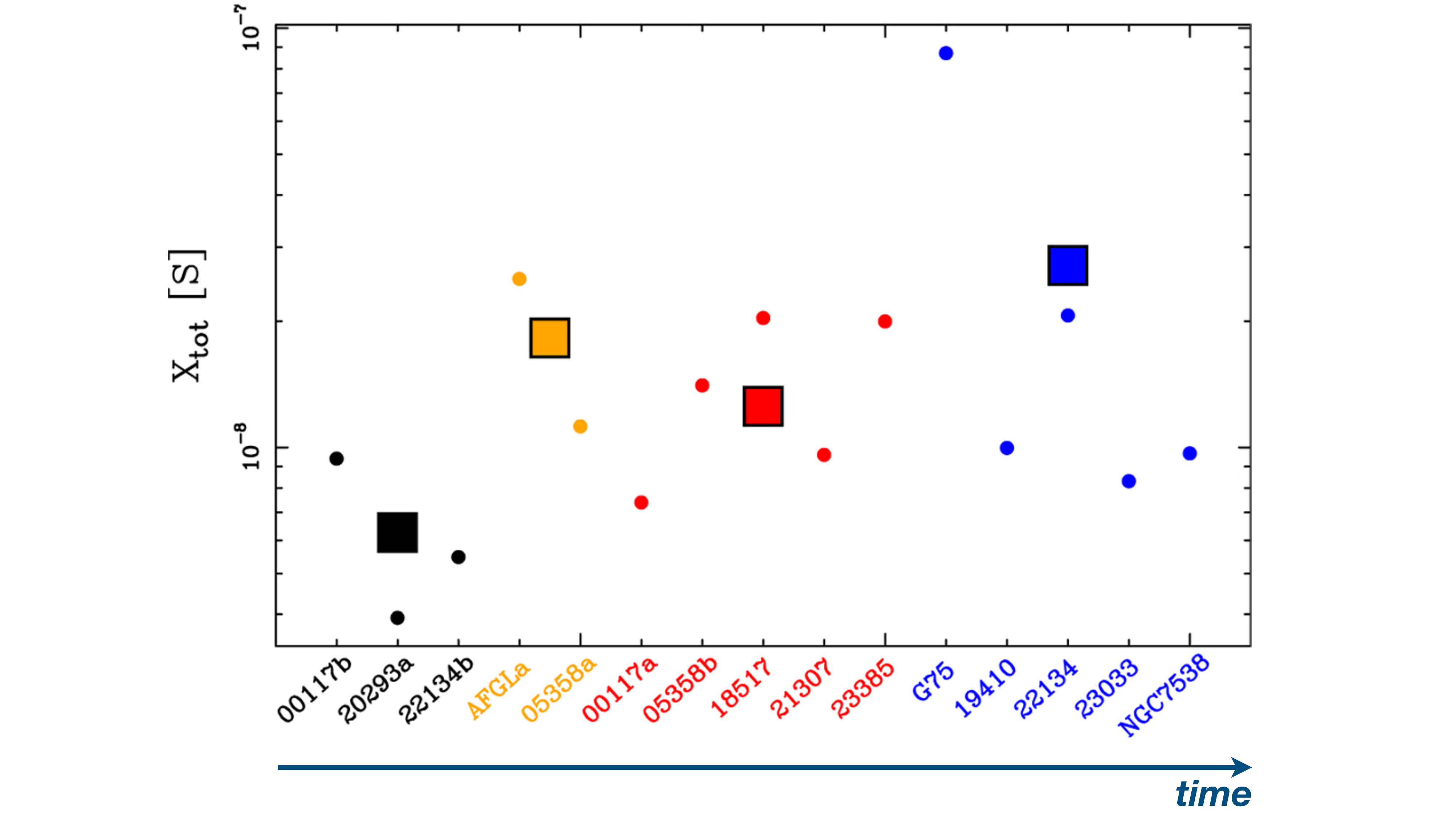}
\caption{Sum of the molecular fractional abundances calculated towards S-bearing species ($X_{\rm tot}$[S]) in each source of the evolutionary sample studied in \citet{fontani23}. 
The source names are indicated on the x-axis. 
The small symbols represent the total molecular abundances in each source, and the large symbols are the average values calculated in each evolutionary group: cold HMSCs (in black), warm HMSCs (in orange), defined as massive starless cores with kinetic temperature $\leq 20$~K and $> 20$~K, respectively, HMPOs (in red), and \UCs(in blue). 
The highest value of $X_{\rm tot}$[S] is $\sim 10^{-7}$ (measured towards the \UC\ region G75-core), but it is two orders of magnitude smaller than the S elemental abundance \citep[$\sim 1.7 \times 10^{-5}$,][]{lodders03}.
From \citet{fontani23}.}
\label{fig:sulphur}
\end{figure}

Some sulphur-bearing molecules, in particular SO$_2$ and H$_2$S, are also typical tracers of protostellar outflows and jets, as indicated by the correlation with SiO both in abundance \citep[e.g.][]{luo24} and in spatial emission \citep[e.g.][]{fontani24}.
Despite a long history of study, sulphur chemistry in the interstellar medium still faces the problem that its main reservoir is unknown.
Observations of gaseous S-bearing molecules \citep{woods15,vastel18,riviere20,bouscasse22,fontani23} indicate a sulphur abundance in the molecular gas of approximately $\sim 1-10$ percent of the elemental cosmic value ($1.73 \times 10^{-5}$, Lodders~\citeyear{lodders03}), with a strong dependence on the environment \citep{fuente23}.
However, observations of S-bearing molecules on ices provide abundances that are too low to solve the problem of the depletion seen in the gas \citep[e.g.][]{boogert15,mcclure23}.
Therefore, any attempt to study the chemical evolution of S-bearing molecules is challenging.
From a pure observational point of view, several S-bearing species were proposed as evolutionary indicators in surveys of high-mass star-forming cores.
\citet{fontani23} observed simple S-bearing species towards a limited sample of HMSCs, HMPOs, and \UC s,  previously studied in various isotopic fractions \citep{fontani11,fontani14,fontani15,fontani15b,colzi18a}.
They propose that species as CS, CCS, HCS$^+$, and NS, trace preferentially quiescent (i.e. less evolved) gas, while species as OCS, and SO$_2$
trace more turbulent (i.e. more evolved) material.
They investigated how the molecular abundances vary as a function of the evolutionary indicators, in particular \Tk\ and the $L/M$ ratio, and found that the best positive correlations are found for SO and SO$_2$, and in general for oxygen-bearing species \citep[see also][]{martinez24}, perhaps due to the larger availability of
atomic oxygen with evolution produced by photodissociation of
water \citep[e.g.][]{vandishoeck21}.
The total sulphur abundance in molecules is also found to increase with evolution (Fig.~\ref{fig:sulphur}), likely due to the increasing amount of S that is sputtered
from dust grains owing to the increasing protostellar activity.
In fact, the enhancement is significant especially from the earliest phase, classified in Fontani et al.~\citeyear{fontani23} as cold HMSC with kinetic temperatures $\leq 20$~K, to the later stages, and
the total sulphur gaseous abundance is at most $\sim 10^{-7}$.
A similar trend was found by \citet{tang24}, who computed a total S-bearing molecular abundance attaining at most $6.9 \times 10^{-7}$ in the protostellar phase.
However, these total sulphur molecular abundances are two orders of magnitude lower than the elemental one, confirming that sulphur is highly depleted from the gas phase even in the evolved stages.

%-----------------------------Figure End------------------------------

Several molecular abundance ratios of sulphurated species have been investigated to find trends with evolution. 
One of the most promising candidates is probably the SO/SO$_2$ ratio
\citep[e.g.][]{herpin09,fontani23,martinez24}.
This ratio was proposed to decrease with time in massive cores \citep{wakelam11}.
The SO$_2$ abundance can increase with respect to that of SO either via surface chemistry (by oxygenation of SO), or via the gas-phase barrierless reaction \citep{vastel18}:
\begin{equation}
    {\rm SO + OH \rightarrow SO_2 + H}\;,
\end{equation}
which is favoured in the late warm stages by the enhanced presence of OH in the gas.
This decrease with evolution is confirmed in the single-dish observational surveys of \citet{fontani23} and \citet{martinez24}, but not in \citet{herpin09}.
Other ratios proposed as possible evolutionary indicators involve CS, such as SO/CS, OCS/CS, and CS/H$_2$S \citep{herpin09,li15,elakel22,fontani23}.
In particular, \citet{elakel22} proposed the SO/CS ratio as a suitable tool to distinguish between warm (SO/CS$> 1$) and cold (SO/CS$< 1$) chemistry within the same source.
However, we must bear in mind that in many of these works the sample size is small \citep[e.g.][]{herpin09,fontani23} and the dispersion of values is large, and therefore these proposed chemical clocks must be taken with caution.
A particularly interesting case is the SO$_2$/OCS ratio, since SO$_2$ and OCS are the only S-bearing species detected so far in interstellar ices \citep[e.g.][]{palumbo97,boogert22,mcclure23,rocha24}.
Both molecules are commonly detected in the gas, but while SO$_2$ is a well-known outflow tracer and it is believed to have an important gas-phase formation route \citep[][,see above]{vastel18}, OCS is more likely a surface chemistry product when CO ice is abundant on grains \citep[e.g.][]{ferrante08,elakel22} and is not commonly found in outflows \citep[e.g.][]{drozdovskaya18}.
In the gas, \citet{herpin09} found a SO$_2$/OCS ratio that increases with evolution. 
\citet{santos24} indicate that OCS is more strongly linked to \METH, a pure surface chemistry product, than SO$_2$. 
This finding is in agreement with models \citep{vew18} that predict a reprocessing of SO$_2$ upon desorption more efficient than for OCS, and an increase of the SO$_2$/OCS ratio at $T\sim 100$~K on core timescales $10^{4}-10^{6}$~yrs.
The negligible contribution of gas-phase processes in the formation of OCS is corroborated by gas-phase models, which are not able to reproduce observed OCS abundances \citep{loison12}.

Finally, \cite{li17} report an enhancement of c–C$_3$H$_2$ and deuterated species such as NH$_2$D, DCO$^+$, and DCN toward shocked regions, as suggested by the presence of SiO and/or SO emission, in their spectral line mapping observations towards four massive star-forming regions (Cepheus A, DR21S, S76E, and G34.26+0.15). 

\subsubsection{Masers}
\label{maser}

Although strictly speaking maser lines are not chemical tracers, they are excellent signposts of high-mass star forming regions \citep[e.g.][]{ellingsen10}
and were proposed as evolutionary indicators because of their quick variations with the change in local physical properties.
Therefore, without going into the details of the mechanisms responsible for the formation of interstellar masers \citep[see, e.g.,][]{elitzur89}, we mention some studies that highlighted the appearance of particular maser lines along the evolution of the high-mass star formation process.
Masers of methanol, water, and OH are typically observed towards both the HMPO and \UC\ stages.
Methanol is likely the species towards which the highest number of maser lines were detected.
\citet{menten91} suggested to classify them into class I and class II types based on their position: class I masers are thought to be offset from bright infrared sources and \UC\ regions, while class II masers are spatially associated with them in projection.
For this reason, class I masers are believed to be collisionally pumped, while class II masers radiatively pumped.
In both cases, they tend to appear very early because of the high column density of \METH\ they need to be excited, but differences were highlighted between specific lines.
A tentative evolutionary sequence was proposed by \citet{breen10}, based on observations of the two class II methanol masers at 6.7~GHz and 12.2~GHz, towards dusty clumps with or without an embedded \UC\ region detectable in the radio continuum.
They proposed that the less evolved clumps are associated only with the 6.7~GHz maser line, and then the 12.2~GHz line appears when the \METH\ column density decreases significantly, that is after the formation of the \UC\ region.
Several studies \citep[e.g.]{breen10,beltran18b} proposed a coarse evolutionary sequence in which class I \METH\ masers appear in the earliest protostellar phases, followed by class II \METH\ masers and then \WAT\ masers.
Water masers would tend to appear later because they need significant outflows or winds to develop well.
The latest stages are characterised by the appearance of OH masers \citep[e.g.][]{gel99,breen18}, a molecule believed to be abundantly produced via the photodissociation of water by UV photons.
However, other studies suggest that there are no significant evolutionary differences between class I and II \METH\ masers \citep[e.g.][]{fontani10}, and that \WAT\ masers tend to appear even before class II \METH\ masers \citep{ladeyschikov22,yang23}.
Therefore, it is very difficult to define an evolutionary sequence only with masers because the different types can be found almost in all  evolutionary stages \citep[e.g.][]{yang20}.

\subsubsection{Summary, open questions, and future prospects}

The results obtained so far from source surveys towards specific lines or species indicate that the clearest evolutionary indicators are deuterated fractions of a few molecules (e.g. \H, H$_3^+$).
Tentative trends are proposed for some carbon-chain molecules, COMs, and sulphur-bearing species, as well as for some abundance ratios (e.g. \H/HCO$^+$, \H/CCS, SO/SO$_2$), although with remarkable diversity among different surveys.

The main difficulty in studies of this kind is that source surveys are never perfectly homogeneous.
Sources in similar evolutionary stages can have significantly different physical properties, making it difficult to disentangle the effect of individual parameters on the observed chemical abundances.
Even when objects are coeval, variations in temperature, density, irradiation from cosmic rays and UV photons, and/or initial elemental abundances, can produce markedly different observational chemical signatures. 
Environmental effects, such as isolated versus crowded regions, can also affect the chemical composition especially in the external envelope of star-forming cores.
Moreover, source surveys often contain objects at different distances, which implies that the observed emission may arise from significantly different linear scales.
To solve these problems, observations of objects with both homogeneous heliocentric distances and very similar (and well-measured) physical parameters are critical to  constrain the parameter space, and help isolating the effect of time on the observed molecular abundances.

The results summarised above concern solely the evolution of gaseous species.
Therefore, a relevant open question that still needs to be addressed is whether the molecular composition of iced dust grain mantles also evolves with time.
During the star-formation process, especially in the early cold phases, atoms and molecules freeze-out on dust grains forming ice mantles, in which surface chemistry contributes to the molecular complexity. 
Because of the continuous adsorption on and sublimation from the ice mantles, the mutual exchange between gaseous and iced species plays an essential role in regulating the evolution of molecular abundances.
Therefore, studying both the composition of gas and ices is critical to have a complete picture of molecular formation processes, since many species can form via both gas-phase and surface chemistry processes \citep[e.g.,][]{cec12,ceccarelli23}.

Ice mantles can be identified and studied through vibrational modes observed in absorption against near- and mid-infrared continuum background sources.
The James Webb Space Telescope (JWST) is the best facility currently available to perform such comparative studies of gas and ices.
Observations from JWST programs like Ice Age
\citep{mcclure23} and JOYS \citep{vandishoeck25}
have begun to give insights into the composition of the two phases.
Both programs have identified a wide range of iced molecules including simple and complex organic molecules. 
However, none of these programs are focussed on high-mass objects, nor on their evolution.
Future JWST surveys of high-mass cores in well-defined evolutionary stages will play a fundamental role in attacking this problem.
In particular, sources with well-constrained physical parameters and gas-phase abundances of key species will be eminently suitable to unveil the role of evolution in shaping the molecular composition of ice grain mantles.

\section{Theory}
\label{theory}
In this Section, we will review theoretical studies devoted to the chemistry in high-mass 
star-forming regions. Cold chemistry in HMSCs is followed by warm-up, and warm chemistry in HMPOs, 
and evolves into photon-dominated chemistry in HII regions. Since timescales of dynamical evolution of 
high-mass protostars that also have complex morphology are short~\citep[][]{ZinneckerYorke07}, and 
comparable to or even shorter than those of chemical evolution, chemistry appears to be linked to dynamics 
more tightly than in low-mass protostars. Although, as stated in Sect.~1, distinction between 
the evolutionary stages is not always clear, it is nevertheless possible to identify chemical species 
and processes related to the phases proposed according to Fig.~1. 

Chemical modelling is a powerful tool not only for exploring chemical abundances, but also for estimating key physical properties of star-forming regions, such as gas density, dust and gas temperatures, cosmic-ray ionisation rates, ionisation fractions, and dynamical evolution timescales.
%The ages of protostellar objects can be estimated indirectly through the so-called ``chemical age'', which is a moment in time-dependent astrochemical model corresponding to the best agreement between modelled and observed abundances of chemical species. 
As can be seen, some studies employ a simplistic physical setup, but explore chemical processes in great detail. Other studies are focussed on construction of complex models that combine chemistry, dynamics, and radiative transfer. In such studies, chemistry often serves as a tool for investigating physical characteristics of massive star-forming regions.
%Thus, the Section will be divided in three parts reflecting the major stages .
\subsection{Chemistry in HMSCs and IRDCs} 
%chemical modelling of cold massive cores will be reviewed and compared to that of low-mass cores. It will be shown that somewhat higher dust temperature in HMSCs than in low-mass prestellar cores ($\sim$20--25~K vs. $\sim$10~K) can produce different abundances of both some simple and complex species (Vasyunina et al., 2012, 2014). The possibility of the formation and detection of COMs in IRDCs will be discussed in the context of new experimental data on non-thermal desorption of complex molecules (Dartois et al., 2019). Inferences on dynamical history of high-mass starless cores from theoretical studies of the evolution of deuterium fraction with chemodynamical models will be discussed (Kong et al.~2015, 2016). 
%Finally, an attempt to identify the high-mass starless core on the earliest stage of evolution using chemical modelling will be discussed (Cyganowski et al., ApJ, 2014).
High-mass starless cores are often embedded in infrared-dark clouds (IRDCs). Those are characterised by low temperatures and high densities, and likely represent the earliest stages of development of high-mass protostars. Although physical conditions in HMSCs and IRDCs are mostly similar to that in low-mass prestellar cores, there are indications that some differences may exist, including somewhat higher temperature ($\sim$15-20~K in IRDCs vs. $\le$10~K in low-mass prestellar cores). Given the exponential dependence of grain surface-related chemical processes on temperature, such small temperature difference may impact observed chemical composition of IRDCs. This possibility was explored with chemical models.

\citet[][]{Vasyunina_ea12} studied chemical evolution of infrared dark clouds (IRDCs) utilizing a pseudo-time-dependent, zero-dimensional chemical modelling approach. Their simulations considered three chemical networks: a purely gas-phase network, a network combining gas-phase reactions with accretion and desorption processes, and a complete gas-grain network incorporating detailed grain-surface chemistry. They specifically analysed two representative IRDCs: the colder IRDC013.90-1 (temperatures $\sim$15~K) and the warmer IRDC321.73-1 (temperature $\sim$25~K). The authors compared model-predicted abundances against observed abundances for several simple molecular species, including N$_2$H$^+$, HC$_3$N, HNC, HCO$^+$, HCN, C$_2$H, NH$_3$, and CS. Chemistry of those species is mainly governed by gas-phase chemical reactions. Interestingly, the complete gas-grain chemical network consistently demonstrated the best agreement with observational data. It was found that at the dust temperature range between 20~K and 30~K grain surface chemistry impacts abundances of certain simple species in the gas phase, such as N$_2$H$^+$. At these intermediate temperatures, grain-surface chemical reactions becomes notably efficient in removing gaseous CO. At dust temperature above 20~K, the CO molecules do not freeze-out efficiently. They rather involved in an accretion-desorption cycle due to collisions of CO molecules with interstellar grains. \citet[][]{Vasyunina_ea12} showed that despite low residence time on grain surface, CO participate in surface reactions with OH and S, forming CO$_2$ and OCS (prefix (gr) denotes reactions on grains):
\begin{equation}
{\rm CO_{(gr)} + OH_{(gr)} \rightarrow CO_{2(gr)} + H_{(gr)} } 
\end{equation}
\begin{equation}
{\rm CO_{(gr)} + S_{(gr)} \rightarrow OCS_{(gr)}}
\end{equation}
Those reactions effectively decrease the abundance of the gas-phase CO at the temperature range 20--30~K at which CO freeze-out is not efficient. This depletion of gas-phase CO significantly decreases its ability to destroy N$_2$H$^+$ via the reaction:
\begin{equation}
{\rm N_2H^+ + CO \rightarrow HCO^+ + N_2}
\end{equation}
Consequently, the abundances of N$_2$H$^+$ become notably elevated in environments within this temperature window, where gaseous N$_2$ that has similar binding energy to that of CO but is inert in surface chemistry, remains abundant, yet CO is efficiently depleted onto grains. The N$_2$H$^+$ is formed through the reaction:
\begin{equation}
{\rm H_3^+ + N_2 \rightarrow N_2H^+ + H_2    }
\end{equation}
The described chemical interactions illustrate how the temperature range of 20–-30~K is crucial for distinguishing between various types of star-forming regions. Warm IRDCs in this temperature regime demonstrate noticeably different chemical behavior compared to colder low-mass starless cores, where extensive freeze-out dominates, and warmer massive protostellar objects, where higher temperatures keep CO primarily in the gas phase.

\citet[][]{vasyunina14} investigated how complex organic molecules form in cold, dense Infrared Dark Clouds (IRDCs). They surveyed 43 IRDCs both in Northern and Southern hemispheres. Estimates of dust temperature in this set of IRDCs vary in a range of 15--25~K. Those temperatures were thought to be too cold for efficient COM formation, yet observations revealed species such as acetaldehyde (CH$_3$CHO), methyl formate (CH$_3$OCHO), and dimethyl ether (CH$_3$OCH$_3$). 
%The authors detected eight molecules across northern and southern IRDCs, including CO, H$_2$CO, HNCO, CH$_3$CCH, CH$_3$OH, CH$_3$CHO, CH$_3$OCHO, and CH$_3$OCH$_3$.
Estimated abundances for CH$_3$CHO, CH$_3$OCHO, and CH$_3$OCH$_3$ in IRDCs were reported for the first time, typically ranging from 10$^{-10}$ to 10$^{-8}$. COM abundances in IRDCs were found to be higher than in low-mass prestellar cores but lower than in evolved hot cores or high-mass protostellar objects (HMPOs). In the observed set of cores, the highest gas temperature was found in IRDC028.34+0.06: \citet[][]{sanhueza_ea12} estimated it as ~30~K. This is the only core in the set where methyl formate (CH$_3$OCHO) was detected. 

To understand how the COMs typically found in hot cores with temperatures higher than 100~K can be formed in much colder environments of IRDCs, using the astrochemical model, the authors tested several scenarios for IRDC028.34+0.06. They found that models with fixed temperatures (15--120~K) and density (10$^5$ cm$^{-3}$) could not reproduce observed COM abundances, underestimating them by 2--10~orders of magnitude. 
%Grain-surface reactions that lead to the formation of COMs are inefficient at 15--20~K due to negligible mobility of reactant radicals. At higher temperatures of $\ge$25~K, radicals on surface are generally mobile, however their abundances are too low. This is because the radicals are formed via photolysis from species such as methanol and formaldehyde. In turn, those species are products of CO hydrogenation that requires temperatures near 10~K. 
A sort of a warm-up scenario which was originally proposed by \citet[][]{GarrodHerbst06} for hot cores was more successful in reproducing the abundances of COMs in IRDC028.34+0.06. The model considers two stages. During the cold stage, chemistry develops during 10$^6$~years at 10~K and constant gas density of 10$^5$~cm$^{-3}$. In the warm-up phase, the temperature of gas and dust in the model increases from 10~K to 30~K over 6.5$\times$10$^4$~years. This modest warm-up is sufficient to make radicals on grain surface mobile and to form COMs. In warm-up models for hot cores, temperature eventually increases to 200~K or higher resulting in complete grain mantle evaporation including newly formed COMs. 
In contrast, in the warm-up model applied to IRDC028.34+0.06, the icy grain mantle remains intact as final temperature is only 30~K.
The surface-formed COMs are released to gas at the onset of their formation via reactive desorption. 
In \citet[][]{vasyunina14}, the assumed efficiency of reactive desorption for COMs is $\sim$~1\%. 
Although the assumption of efficient reactive desorption for complex molecules may be doubtful, COMs may be delivered to gas via other non-thermal desorption mechanisms. 
For example, \citet[][]{Dartois_ea19} demonstrated that cosmic-ray-induced sputtering is an efficient non-thermal desorption mechanism for COMs such as methanol (CH$_3$OH) and methyl acetate (CH$_3$COOCH$_3$). 
Unlike VUV photodesorption, which is inefficient for large COMs due to photolysis, cosmic-ray sputtering effectively releases unfragmented molecules. 
The need for an evolutionary chemical model for the explanation of observed molecular content supported the claim that IRDCs are the early stages of massive star formation.

\citet[][]{Entekhabi_ea22} carried out astrochemical modelling of the infrared dark cloud G28.37+00.07. 
The authors aimed to constrain the cosmic-ray ionisation rate (CRIR) and the chemical age in various regions of the IRDC. 
By combining observations of molecular lines such as C$^{18}$O, H$^{13}$CO$^+$, HC$^{18}$O$^+$, and N$_2$H$^+$ with gas-grain astrochemical models, the authors constrain the CRIR and chemical age across ten positions within the cloud that sample various levels of star formation activity. 
The analysis finds typical conditions in the cloud characterised by densities nH$\sim$3$\times$10$^4$-10$^5$~cm$^{-3}$, temperatures around 10-15~K, and CO depletion factors between 3 and 10. 
Results indicate relatively low cosmic-ray ionisation rates ($\zeta\sim$10$^{-18}$~s$^{-1}$ to 10$^{-17}$~s$^{-1}$). 
Notably, no systematic variation in CRIR was observed as a function of star-formation activity in different locations of the cloud. Chemical ages of the regions were generally estimated to be at least several free-fall times ($\sim$3$\times$10$^5$~years), although modelling with additional species like HCN, HNC, HNCO, CH$_3$OH, and H$_2$CO tended to suggest somewhat older ages and higher CRIRs.

A notable work by \citet[][]{Sabatini_ea21} introduce a time-dependent modelling method to estimate evolutionary timescales in massive star formation, improving upon previous static or statistical approaches. 
They combine a simplified isothermal collapse phase with a detailed one-dimensional warm-up phase, incorporating radiative transfer simulations to accurately track temperature, density, and chemical evolution simultaneously. 
The model is applied to a representative sample of massive clumps from the ATLASGAL-TOP100 survey. 
Using chemical tracers such as CH$_3$CCH, CH$_3$CN, H$_2$CO, and CH$_3$OH as chemical clocks, the study successfully reproduces observed molecular abundances, particularly for CH$_3$CCH and CH$_3$CN, though less accurately for H$_2$CO and CH$_3$OH. 
The estimated total duration for massive star formation is approximately 5.2$\times$10$^5$~years, with the individual evolutionary phases clearly distinguished. 
The specific phases were found to last approximately 5$\times$10$^4$~years for the earliest (70-µm weak), 1.2$\times$10$^5$ years for mid-IR weak, 2.4$\times$10$^5$ years for mid-IR bright, and 1.1$\times$10$^5$~years for the most evolved HII-region phase. 
The novelty of this study is primarily in its combination of radiative transfer and chemical modelling in a time-dependent manner, moving beyond previous static density and temperature assumptions. 
The results indicate that molecular abundances significantly depend on the thermal evolution of the clumps, notably displaying clear evaporation fronts within the modelled clumps.

Deuterated molecules can serve as another powerful diagnostic tool for the earliest evolutionary stages of both low- and high-mass star forming regions. \citet[][]{Kong_ea15} investigated how rapidly deuterium enrichment builds up in the dense, cold gas of star‑forming cores by modelling the ratio [N$_2$D$^+$]/[N$_2$H$^+$] with a new,  expanded chemical network that keeps track of deuterium chemistry and the spin states of H$_2$ and H$_3^+$ isotopologues. Their network also contained reactions involving H$_3$O$^+$ and its deuterated forms. Using this framework the authors carry out the first uniform exploration of how deuterium fractionation depends on four key physical controls—volume density, temperature, cosmic‑ray ionisation rate and the gas‑phase depletion factor of heavy elements—over ranges that cover both low‑mass and massive starless cores. The simulations show that reaching a deuterium fraction of 0.1 -- the level often observed in both low‑ and high‑mass starless cores —- requires at least several local free‑fall times. Because these chemical timescales match the ambipolar‑diffusion time rather than the dynamical free‑fall time, the authors argue that significant magnetic support must be retarding collapse in highly deuterated cores.  
%Application of the model to the well‑studied low-mass pre‑stellar core L1544 yields a chemical age five to ten times its present free‑fall time, reinforcing this conclusion. 
The deuterium fraction of 0.1 are favoured by densities above a few $\times$10$^4$~cm$^{-3}$, temperatures below about 17~K, depletion factors above $\sim$6 and cosmic‑ray ionisation rates below $\sim$6$\times$10$^{-17}$~s$^{-1}$; 

\citet[][]{kong16} performed similar study of two massive starless cores, C1-N and C1-S, in the IRDC G028.37+00.07. They used multi-line observations of N$_2$H$^+$ and N$_2$D$^+$ to measure deuterium fractionation and to perform chemodynamical modelling aimed at estimating the collapse rate of the cores. The authors find high deuterium fractions [0.2–0.7], far exceeding the cosmic [D]/[H] ratio, and use these results in conjunction with chemodynamical models to infer the cores’ evolutionary states. The models indicate that such high deuteration levels can only be achieved if the cores are collapsing slowly—at less than one-tenth the free-fall rate—implying near-virial equilibrium likely supported by magnetic fields.

\subsection{Chemistry in HMPOs and hot cores} 
This stage of protostellar development is characterised by the richest chemical composition. In particular, high abundances of complex organic and prebiotic molecules are observed (Sect. 2). Here, we will review recent advances in modelling of hot core chemistry. The development during the last decades of models of hot core chemistry from a basic two-step ``cold collapse -- warm-up'' approach 
%(Garrod \& Herbst, 2006) 
into advanced tools for detailed understanding of chemical evolution 
%(e.g., Choudhury et al., 2015; Bonfand et al., 2019) 
that combines chemical models with radiative transfer, and results of radiation-magnetohydrodynamical 
simulations.
%, will be discussed. 

%The role of less sophisticated 1D models in understanding chemistry of simple species in HMPOs will also be discussed (e.g., Gerner et al., 2014, Gieser et al., 2019), as well as possibility to constrain physical conditions in HMPOs (e.g., Barger \& Garrod, 2020). Understanding important prebiotic phosphorus chemistry in massive star-forming regions with modelling will also be discussed (Fontani et al., 2016, 2019; Rivilla et al., 2016, 2020).

The pivotal role of gradual increase in dust temperature during the development of a low- and high-mass protostars on the formation of complex organic molecules (COMs) in grain-surface chemistry was first recognised in the seminal paper by~\citet[][]{GarrodHerbst06}. 
In hot cores, saturated organic molecules, such as methyl formate (HCOOCH$_3$), formic acid (HCOOH), and dimethyl ether (CH$_3$OCH$_3$) cannot be formed in observed abundances via gas-phase chemistry. 
Therefore, the authors employed a two-stage, time-dependent gas-grain chemical model that simulates both the collapse of a molecular cloud and the subsequent warm-up phase caused by protostellar heating. 
The novelty of this study lies in its detailed treatment of the intermediate warm-up period, during which the temperature of the gas and dust rises gradually from 10~K to more than 100~K over a time span of a hundred thousand years. 
In contrast to earlier models that focussed solely on cold collapse or hot gas-phase chemistry, this work emphasises the chemistry occurring during this transitional phase, particularly grain-surface reactions involving relatively immobile but reactive heavy radicals such as HCO, CH$_3$O, and OH. 
These radicals are produced via photolysis of water and methanol ices. 
%Reduced due to the elevated grain temperatures presence of atomic hydrogen on surface allows the radical to avoid hydrogenation, and instead react with each other. 
\citet[][]{GarrodHerbst06} showed that the formation of methyl formate, formic acid, and dimethyl ether is significantly enhanced by these surface reactions at temperatures between 20~K and 40~K via  radical-radical grain-surface reactions such as:
\begin{equation}
    {\rm HCO~+~CH_3O~\rightarrow~HCOOCH_3}\;,
\end{equation}
\begin{equation}
 {\rm CH_3~+~CH_3O~\rightarrow~CH_3OCH_3}\;, 
\end{equation}
and 
\begin{equation}
   {\rm HCO~+~OH~\rightarrow~HCOOH}\;, 
\end{equation}
originally proposed by \citet[][]{AllenRobinson77}. 
At these temperatures, atomic hydrogen (H) evaporates from grain surfaces, freeing radicals such as OH and HCO from hydrogenation, which then react to produce complex organic molecules. 
Further warm-up beyond the sublimation temperatures of COMs, release them to the gas phase. 
It was also demonstrated that certain gas-phase routes previously considered inefficient become important at lower intermediate temperatures after evaporation of large amounts of reactants from grains, further enhancing molecular complexity. 

\citet[][]{GarrodHerbst06} introduced an important concept relevant for both low-mass and high-mass star formation that links physical development of a protostar with its chemical complexity. 
The key role of radical-radical reactions and photolysis in icy mantles of dust grains in the formation of complex organic and prebiotic molecules was revealed. 
The warm-up model was further expanded on a large number of complex organic and prebiotic species in the following papers by~\citet[][]{garrod08} and \citet[][]{garrod13}.

\citet[][]{Choudhury_ea15} extended a simplistic physical models of first warm-up works into a self-consistent modelling framework to simulate the chemical and spectral evolution of hot molecular cores during high-mass star formation. 
The authors coupled a gas-grain chemical code \textit{Saptarsy} with radiative transfer calculations done using \textit{RADMC-3D} code to track the spatio-temporal evolution of key complex organic molecules (COMs) such as CH$_3$OH, C$_2$H$_5$OH, CH$_3$OCH$_3$, and HCOOCH$_3$. 
They explore how variations in density structure, protostellar bolometric luminosity evolution, and cosmic-ray ionisation rates affect COM formation, desorption, and emission. 
Synthetic (sub-)mm spectra and emission maps are generated under LTE conditions to enable comparison with observations. 
Unlike previous studies that used static or simplified warm-up prescriptions, \citet[][]{Choudhury_ea15} considered dynamic and spatially-resolved treatment of physical conditions, integrating time-dependent temperature structures derived from protostellar  bolometric luminosity models with detailed gas-grain chemistry. It is also one of the first studies to simulate the spectral evolution of COMs in high-mass cores. \citet[][]{Choudhury_ea15} reported modelled COM abundances (e.g., CH$_3$OH $\sim$10$^{-6}$–10$^{-7}$, HCOOCH$_3$ $\sim$10$^{-7}$–10$^{-8}$) to be consistent with observed values, especially for higher cosmic-ray ionisation rates. Simulated spectra reproduce observed trends, including increasing line density and intensity with time, and predict that the spatial extent of emission from more weakly bound molecules (e.g., CH$_3$OCH$_3$) exceeds that of more strongly bound species like CH$_3$OH—offering a potential observational diagnostic of desorption energies.

\citet[][]{Gieser_ea19} used the MUSCLE (MUlti Stage CLoud codE) framework, which combines static 1D physical structure of a modelled core with a time-
dependent gas-grain chemical model ALCHEMIC~\citep[][]{Semenov_ea10} to model AFGL 2591 VLA 3 hot core. By comparing modelled and observed column densities, the authors find that the abundances of 10 out of 14 species are reproduced at a chemical age of ~21,100 years. While observations suggest a core density power-law index of 1.7, the model prefers a shallower gradient ($<$1.4), indicating limitations of the 1D assumption. This study improves on earlier work by integrating high-resolution molecular data with detailed chemical modelling tailored to a specific source, offering a physically grounded estimate of chemical age and structure in a high-mass protostar.

\citet[][]{Bonfand_ea19} combined a warm-up chemical model with a detailed radiative transfer calculations and the results of radiation-magnetohydrodynamic simulations, and applied it to four molecular cores (N2, N3, N4, N5) in Sagittarius B2(N) object close to the Galactic center. The ALMA data from the EMoCA spectral line survey~\citep[][]{belloche16} has been used to benchmark models against observational data. The analysis reveals distinct chemical compositions among the sources: cores N3 and N5 share similar abundances relative to methanol, while the core N2 differs significantly, particularly in its higher abundance of ethyl cyanide (C$_2$H$_5$CN) and formamide (NH$_2$CHO). Chemical modelling with the MAGICKAL code~\citep[][]{garrod08}, coupled with radiative transfer calculations (RADMC-3D) and radiation-magnetohydrodynamic simulations, demonstrated that the production of complex organic molecules (COMs) is highly sensitive to environmental conditions. A cosmic-ray ionisation rate of 7×10$^{-16}$ s$^{-1}$ (50 times the standard interstellar value) best reproduced the observed abundances of ten COMs relative to methanol. This elevated rate accounts for the Galactic center environment while still allowing efficient COM formation. The models further constrained the minimum dust temperature during the prestellar phase in the Galactic center region to $<$25 K, with 15~K enabling efficient COM formation on grains through radical addition reactions. It was found that higher temperatures ($>$25 K) suppress crucial grain-surface chemistry, particularly for cyanides and oxygen-bearing species like methyl formate (CH$_3$OCHO) and dimethyl ether (CH$_3$OCH$_3$). The authors stress that the chemical differentiation between sources, especially the anomalous composition of N2, underscores the importance of individual physical histories. 
A combined hydrodynamic and gas-grain chemical model of a hot core is also presented in~\citet[][]{Barger_ea21}. In that work, gas-grain chemical model is coupled with physical evolution of the core computed using a radiation hydrodynamics model. 
Interestingly, despite employing an advanced physical model of a hot core, the model exhibited somewhat limited agreement with observational data on Sgr B2(N2).

\citet[][]{BargerGarrod20} explored the influence of cosmic-ray ionisation rates (CRIR) and warm-up timescales on the chemical evolution of massive hot cores. 
Using the MAGICKAL code, the authors simulated a grid of 81 models varying in both CRIR and warm-up timescale, and incorporated LTE radiative transfer to produce simulated molecular emission spectra. 
These are then compared against observed data for four well-studied hot cores: NGC 6334 IRS 1, NGC 7538 IRS 1, W3(H$_2$O), and W33A.
The results suggest that elevated cosmic-ray ionisation rates — order of magnitude higher than the canonical value of 1.3$\times$10$^{-17}$~s$^{-1}$ commonly used in previous models, as well as relatively rapid warm-up timescales ($\sim$10$^4$~years) are necessary to reproduce the observed molecular abundances.
Cosmic rays drive radical production via photodissociation and thereby enable radical-radical reactions on grain surfaces, but also contribute to gas-phase destruction of COMs through ion–molecule reactions. 
Interestingly, best-fit CRIR is higher than that estimated by \citet[][]{Entekhabi_ea22} for IRDCs (see Section~3.1).

As it can be seen, the development of models is headed towards creating self-consistent packages that combine chemistry with radiative transfer and hydrodynamics. 
However, models of chemistry themselves also undergone significant improvement, directed towards inclusion of new complex molecules, isomers and new mechanisms of solid-state reactivity. 
Several works related to the EMoCA line survey of Sgr B2(N2) hot core~\citep[][]{belloche16} are the examples of such approach.

Following the first discovery of a branched molecule in the ISM, iso-propyl cyanide in Sgr B2(N)~\citep[][]{Belloche_ea14}, \citet[][]{Garrod_ea17} used the three-phase chemical kinetics model MAGICKAL to simulate the formation of branched carbon-chain molecules in the hot-core source Sgr B2(N2). The updated chemical network explicitly treated radical isomerism and incorporated key grain-surface reactions, particularly the addition of the CN radical to unsaturated hydrocarbons with low activation energy barriers. The modeling identified the dominant formation mechanism for branched isomers. The simulations predict that branching becomes more pronounced with molecular size. The results highlight that efficient CN-addition channels enhance the production of straight-chain nitriles, suggesting that other molecular families without such low-barrier pathways may exhibit even greater degrees of branching. \citet[][]{Willis_ea20} introduced a single-stage ``collapse-warmup'' chemical model of isocyanides formation in Sgr B2(N). A significantly expanded chemical network for the first time included new isocyanides like CH$_3$NC, HCCNC, C$_2$H$_5$NC and C$_2$H$_3$NC, with reactions often inferred from their cyanide analogs. A key finding is the critical importance of using an enhanced and extinction-dependent cosmic-ray ionization rate ($\zeta$) for the reproducing of the CH$_3$NC:CH$_3$CN ratio.

In a recent comprehensive paper by \citet[][]{Garrod_ea22}, a major revision of the MAGICKAL gas–grain chemical kinetics code aimed at explaining the origin of complex organic molecules (COMs) in hot molecular cores is presented. 
They extend the standard diffusive three‑phase framework by embedding non-diffusive mechanisms of surface reactivity first introduced in \citet[][]{GarrodPauly11} and later generalized in \citet[][]{JinGarrod20} to Eley–Rideal encounters, photodissociation‑induced reactions, three‑body (follow‑on) reactions and its excited‑formation variant.
In this way, radicals and stable species can react immediately after they come into contact on grain surfaces or inside the ice mantle without waiting for thermal diffusion.
This is coupled to other upgrades, including revised treatment of diffusion and desorption barriers, new proton‑transfer reactions with ammonia that prolong gas‑phase COMs lifetimes, and an expanded network of reactions. 
Using a standard collapse followed by fast, medium and slow warm‑up tracks, the authors show that nondiffusive chemistry moves the production of many COMs from the classical 20–-50~K diffusive regime to much earlier times and to dust temperatures as low as 10 K. 
Significant fractions of molecules such as methyl formate, dimethyl ether, ethanol and even glycine are synthesised while the ice mantle is still growing, either as by‑products of CO hydrogenation or by cold UV photolysis of nascent ices. 
Additional bursts of synthesis occur when water and other strongly bound ices desorb at 120--160~K, liberating trapped radicals that react on hot surfaces. 
Because ammonia efficiently neutralises protonated COMs in the gas, the modelled abundances remain high long after ice desorption. 
The model simultaneously reproduces the gas‑phase abundances of dozens of COMs across both low‑ and high‑mass protostars, including previously problematic methyl formate~:~glycolaldehyde~:~acetic acid ratio. 

Although prebiotic importance is usually attributed to carbon-bearing complex organic molecules, phosphorus (P), which is synthesised in massive stars, is another crucial element for development of life on Earth. It is part of DNA and RNA molecules, phospholipids, and the adenosine triphosphate molecule. However, chemistry of phosphorus-bearing molecules in star-forming regions is not well understood. Nevertheless, several studies of P-bearing molecules were published during the last decade.

\citet[][]{fontani16} presented detections of phosphorus nitride (PN) in a sample of massive dense cores at different stages of high-mass star formation, ranging from starless cores to protostellar objects and ultracompact HII regions. Interestingly, for the first time PN was found in relatively cold (20--60~K) and quiescent environments, in contrast to previous studies that associated PN predominantly with hot ($\ge$100~K) and turbulent regions influenced by shocks. The authors employed a time-dependent chemical model MONACO based on the work of \citet[][]{VasyuninHerbst13} to understand the formation of detected PN. The model comprises a two-phase evolutionary scenario: an initial cold phase (10 K, n(H$_2$) = 10$^4$ cm$^{-3}$) that lasts 1 Myr, followed by a warm-up phase (final temperatures are T = 40, 50 or 60 K, n(H$_2$) = 10$^5$~cm$^{-3}$) that also lasts 1 Myr. This framework was used to compute the abundance ratios of several species (e.g., CN/PN, CH$_3$OH/PN, N$_2$H$^+$/PN, and HNC/PN), which were then compared with observationally derived ratios. The models reproduced the observed abundance ratios for most species within the expected timescales and temperatures, except for CN/PN, which was likely affected by differing spatial distributions of the molecules. Thus, the possibility of formation of observed abundance of PN in a relatively cold quiescent gas via gas-phase chemistry was confirmed.

\citet[][]{rivilla16} presented first detection of another key phosphorus-bearing prebiotic molecule, phosphorus monoxide (PO) in two massive star-forming regions, W51 e1/e2 and W3(OH). The observational results are supplemented by chemical modelling that clarifies the formation of PO alongside phosphorus nitride (PN). The modelling setup is similar to that utilised in Fontani et al. (2016) with the exception final temperature of the warm-up phase, which is equal to 200~K. The modelling indicated that both PO and PN shall be formed via gas-phase ion–molecule and neutral–neutral reactions already during the cold phase of protostellar development and freeze to grains. The observed PO and PN are likely released to the gas phase from grains at the onset of protostellar warm-up at $\sim$35~K. The observed abundance of PO $\sim$10$^{-10}$ to H$_2$ as well as PO/PN abundance ration of 1.8--3 are successfully reproduced by the model. Interestingly, a relatively high initial abundance of atomic phosphorus of 5$\times$10$^{-9}$, 25 times higher than the ``low metal'' P-abundance typically used in dark cloud chemical models is required to reproduce observations. In a later work by \citet[][]{rivilla20}, an alternative scenario of PO formation in shocked environments is proposed. The scenario implies the formation of PO in gas-phase photochemistry of phosphine (PH$_3$). The latter is previously formed on dust grains via hydrogenation of P atoms and then released to the gas due to shocks.

\subsection{Chemistry in HC and \UC\ regions} 
These regions follow hot cores, and represent the final stage of the development of a massive protostar~\citep[][]{Churchwell02}. At the same time, they have a strong impact on their surrounding molecular clouds via feedback mechanisms \citep[][]{giannetti_ea12, Moscadelli_ea18}, and thus in setting the initial conditions for the next generation of forming stars. Complex physical and chemical evolution occurs there on small spatial scales which makes its study challenging.
%Models of ultra-compact and more developed \HII\ regions will be reviewed (e.g., Stephan et al., 2018) with emphasis on molecular tracers and specific photon-dominated chemistry of hydrocarbons that develops there (Pilleri et al., 2013).
% \textbf{Shall we go here up to developed HII regions?? If so, one can also mention chemodynamical models of expanding HII regions (e.g., \cite{Kirsanova_ea19})}

\citet[][]{Pilleri_ea13} investigated the spatial distribution and chemistry of small hydrocarbons, specifically CH, CCH, and c-C$_3$H$_2$, in the vicinity of the ultra-compact HII region Monoceros R2 (Mon~R2), with a focus on photon-dominated regions (PDRs) exposed to high ultraviolet (UV) radiation fields. The modelling work combines extensive observational data from the IRAM 30~m telescope and the Herschel Space Observatory, which are interpreted using both steady-state gas-phase chemical models (via the Meudon PDR code, \citet[][]{LePetit_ea06}) and time-dependent gas-grain chemical models (via UCL\_CHEM code, \citet[][]{Viti_ea04}). The core modelling approach consists of dissecting the Mon~R2 region into distinct physical components: a highly UV-irradiated thin PDR layer (LPDR), a denser, more molecular high-density PDR shell (LHD), and a surrounding low-density molecular envelope (Lenv). For the PDR layers, the Meudon PDR code is used to simulate steady-state gas-phase chemistry under intense UV irradiation (G$_0\sim$5$\times$10$^5$), while for the cooler, shielded envelope, UCL\_CHEM is employed to capture grain surface processes and time-dependent effects, particularly relevant during cloud collapse and subsequent UV illumination. A key finding of the modelling is that gas-phase steady-state chemistry can reproduce the observed abundances of CH and CCH relatively well in the high-UV illuminated PDR, but it fails—by about an order of magnitude—to account for the c-C$_3$H$_2$ abundance. The inclusion of grain-surface chemistry and non-equilibrium effects, modelled through UCL\_CHEM, is essential to reproduce c-C$_3$H$_2$ abundances in the lower-density envelope. The study shows that the N(c-C$_3$H$_2$)/N(CCH) ratio increases from ~0.004 in the envelope to as high as ~0.03 near the PDR peaks, and this ratio correlates spatially with 8$\mu$m polycyclic aromatic hydrocarbon (PAH) emission, suggesting that considered hydrocarbons could be a product of photoprocessing of PAHs.

\citet[][]{Stephan_ea18} investigated hot cores with embedded hypercompact (HC)/ultracompact HII (UCHII) regions with a focus on the chemical structure and evolution of internal photon-dominated regions (PDRs) that surrounds deeply embedded HC and UCHII regions. The aim was to identify specific chemical tracers that can differentiate these evolutionary phases of massive star formation from hot molecular cores (HMCs). The authors used the astrochemical code \textit{Saptarsy} coupled with radiative transfer simulations (RADMC-3D) to compute spatio-temporal evolutions of chemical abundances and to derive synthetic spectra for different atomic and molecular species. It was found that common molecular tracers like C, N$_2$H$^+$, CN, and HCO are not reliable indicators for distinguishing between internal HII regions/PDRs and hot molecular cores because their distributions do not uniquely trace the inner core regions influenced by strong UV fields. On the other hand, atomic species such as C$^+$ and O robustly trace these internal PDRs, but unobservable with existing facilities due to the very narrow size of these internal PDRs (typically less than 100 AU thick).

An attempt to estimate duration of every phase of development of high-mass protostars using astrochemical modelling is made in~\citet[][]{gerner14}. In this work, a comprehensive modelling effort to trace the chemical evolution of high-mass star-forming regions across four evolutionary stages: infrared dark clouds (IRDCs), high-mass protostellar objects (HMPOs), hot molecular cores (HMCs), and ultra-compact HII (UCHII) regions is presented. The MUSCLE framework was utilised to simulate the chemistry within each evolutionary phase. This approach integrates radial variations in density and temperature, and incorporates an advanced chemical network that includes gas-phase and grain-surface processes, as well as cosmic-ray and UV-induced effects. Using the IRAM 30~m telescope, the authors conducted a molecular line survey at 1~mm and 3~mm of 59 regions spanning all evolutionary stages of massive protostars, the authors fit observed molecular column densities with models by varying physical parameters such as central density, temperature, and chemical age. Each evolutionary stage is treated as a distinct model, with physical and chemical properties constrained independently. The iterative modelling yields chemical ages of approximately 10,000 years for the IRDC stage, 60,000 years for HMPOs, 40,000 years for HMCs, and 10,000 years for UCHII regions, aligning with theoretical predictions of a $\sim$10$^5$~years of total duration for massive star formation.

\subsection{Summary, open questions, and future prospects}
Chemical evolution of high-mass star-forming regions exhibit many similarities to their low-mass counterparts. However, shorter timescales of dynamical evolution and stronger radiation fields imply that chemistry in high-mass protostars should be considered as tightly coupled to their dynamics. Chemical modeling is used not only to predict molecular abundances but also to constrain physical properties such as density, temperature, cosmic-ray ionization rates, and evolutionary timescales.

In HMSCs and IRDCs, chemical models reveal that even small temperature differences (e.g., 15-–30~K) significantly affect molecular abundances. 
For instance, at 20-–30~K, CO is efficiently depleted from the gas phase via grain-surface reactions, enhancing N$_2$H$^+$ abundances. 
Complex organic molecules (COMs) are detected in cold IRDCs. 
While early models suggested that a modest warm-up phase and non-thermal desorption mechanisms—such as reactive desorption or cosmic-ray sputtering—are responsible for the formation and release of these molecules into the gas without full mantle evaporation, recent modelling highlight the importance of non-diffusive surface reactivity in the formation of COMs~\citep[e.g.][]{JinGarrod20, Garrod_ea22, Borshcheva_ea25}. Deuterium fractionation, particularly the [N$_2$D$^+$]/[N$_2$H$^+$] ratio, serves as a sensitive chemical clock, indicating slow collapse rates and significant magnetic support in prestellar cores.  

For HMPOs and hot cores, chemical models have evolved from simple two-stage (collapse followed by warm-up) approaches to sophisticated frameworks combining gas-grain chemistry, radiative transfer, and radiation-magnetohydrodynamic simulations. The gradual warm-up (from $\sim$10~K to over 100~K) is critical for forming COMs via radical-radical reactions on grain surfaces. Recent models also incorporate non-diffusive reaction mechanisms, allowing COM formation at temperatures as low as 10~K, and highlight the importance of cosmic-ray-induced chemistry and gas-phase destruction pathways for complex species.

In hyper- and ultra-compact \HII\ regions, chemistry becomes dominated by UV irradiation, resembling photon-dominated regions (PDRs). 
Models highlight the role of atomic lines and recombination lines, and the difficulty of tracing internal PDRs with common molecular tracers. Time-dependent models combining chemistry and radiative transfer provide estimates for the duration of each evolutionary phase, with total massive star formation timescales around 10$^5$~years.

Overall, theoretical advances are moving toward more self-consistent frameworks that integrate chemistry, dynamics, and radiative transfer, while also incorporating new grain-surface reaction mechanisms—such as non-diffusive processes—that allow COM formation at colder temperatures and earlier evolutionary stages. Although analysis of ice composition in some high-mass star-forming regions is already published~\cite[see e.g.][]{Nakibov_ea25}, chemical modelling of the evolution of icy mantles of interstellar grains in high-mass star forming regions is yet to be done comprehensively.

\section{Comparison with other astrophysical environments}
\label{comparison}

In this section, we briefly highlight the main differences and similarities among high-mass star-forming cores and other astrophysical environments from the point of view of the chemical evolution.

\subsection{Low-mass star-forming cores}

As stated in Sect.~\ref{intro}, the low-mass star formation process follows a theoretical scenario significantly different from that of high-mass stars: high-mass star-forming cores are characterised by larger temperatures and densities and evolve in much shorter timescales.
These physical differences are expected to impact the evolution of molecular abundances, and chemical evolutionary studies that compare low- and high-mass star-forming cores are starting to highlight similarities and differences.

Probably, the chemical evolutionary indicator that most closely links low- and high-mass star-forming cores, both from the theoretical and observational point of view, is deuterium fractionation.
The abundance of deuterated molecules, especially \D\ and H$_2$D$^+$,
is found to be greatly enhanced in pre-stellar cores \citep{caselli03,crapsi05,pagani07},
and $R_{\rm D}$ has a clear similar decreasing trend with increasing temperature, and hence protostellar evolution, as in the high-mass case, particularly in \H\ \citep{emprechtinger09,friesen13,ceccarelli14}.
As in the high-mass case, evolution does not seem to play a role in the fractionation of nitrogen, since the \Nfrac\ ratio does not vary significantly with time in low-mass star-forming cores \citep[e.g.][]{desimone18}, except for a depletion of $^{15}$N in \H\ in pre-stellar cores \citep[e.g.][]{bizzocchi13,redaelli18},
which causes \Nfrac\ to attain values $\sim 1000$, never measured so far in high-mass star-forming regions.

Another possible chemical link between low- and high-mass star formation is represented by sulphurated species, COMs, and carbon chains.
Sulphur-bearing molecules are detected at any stage of the low-mass star formation process \citep[e.g.][]{lefloch18}.
\citet{bef03} reported a tentative decrease of the SO/SO$_2$ abundance ratio from the class 0 to the class I protostellar stage, confirmed by \citet{ghosh24}.
This trend is similar to what predicted \citep{wakelam11} and observed \citep[e.g.][]{herpin09,martinez24} in massive cores, although other works show different results.
For example, studying a sample of class I protostars, \citet{legal20} found no anti-correlation between the SO/SO$_2$ abundance ratio and the disk-to-envelope mass ratio, expected to increase with evolution.
A similar result was obtained with ALMA in a sample of class 0 and class I objects in Perseus \citep{artur23}.
The link between the low- and high-mass case is even more uncertain in the evolution of COMs.
While in the high-mass star formation process all COMs keep increasing their abundance with evolution \citep{coletta20}, in low-mass cores the situation is different.
It is well known that warm regions ($T\geq 100$~K) around low-mass protostars are enriched by COMs. 
These compact cores, called hot corinos \citep[e.g.][]{ceccarelli07}, have physical and chemical properties similar to HMCs and are detected both in early (class 0) and evolved (class I) low-mass protostars.
Some observational studies propose an abundance peak of COMs in the class 0 phase followed by a decrease in the class I phase \citep[e.g.][]{bhat23}, due to the massive release in the gas of molecules produced on ice mantles. 
However, similar studies do not highlight clear evolutionary trends, and show that the abundance of COMs do not substantially evolve from the class 0 to the class I protostellar phase \citep[e.g.][]{mercimek22,ceccarelli23}.
Similarly to COMs, observations of carbon chains toward low-mass young stellar objects revealed that these species are as commonly detected as in high-mass protostars \citep{taniguchi24}.
However, their formation around low-mass objects would be favoured by the processes know as 'warm carbon chain chemistry' \citep[WCCC,][]{sey13},
a chemistry initiated by the evaporation of methane from icy grain mantles, but observations of cyanopolyynes towards high-mass star-forming cores indicate that their chemistry occurs differently and requires much higher temperatures \citep{taniguchi21}.

The chemical complexity in low-mass star-forming cores appears to be more affected by local environmental conditions than by evolution \citep[e.g.][]{vangelder20,scibelli24}.
In particular, it is now clear that energetic phenomena associated with high-mass stars, such as high UV irradiation and enhanced cosmic-ray ionisation rates, have a significant impact on the chemistry of some species, like carbon chains and hydrocarbons.
This brings us to the link between the chemical evolution in high-mass star-forming cores and the chemistry inherited by protostellar envelopes forming in high-mass clusters, including the protosolar nebula (Sect.~\ref{intro}).
The prototypical source where this effect is apparent is
OMC-2 FIR4, an intermediate-mass protocluster in Orion north to the Trapezium OB star cluster \citep{chini97}.
The comparison of the high abundances of the \CIII\ and \CV\ cyanopolyynes
\citep{fontani17} and of \cyclic\ \citep{favre18} with the predictions of astrochemical models indicates that OMC-2 FIR4 is irradiated by a Far-UV field $\sim 1000$ times larger than the interstellar one, and associated with a cosmic-ray ionisation rate 
3 orders of magnitude larger than the canonical interstellar value ($\sim 10^{-17}$ s$^{-1}$). 
The dose of energetic particles responsible for this observed high ionisation rate, causing the abundance enhancements in cyanopolyynes and hydrocarbons, would be comparable to that experienced by the young Solar system \citep{ceccarelli14b}.
Therefore, it is tempting to speculate that such high energetic phenomena may have had an important impact in the budget of pre-biotic material inherited from the protosolar nebula.
Indeed, cyanopolyynes were detected in on Titan’s atmosphere \citep[e.g.][]{vuitton07} and comets \citep[e.g.][]{mec11}, the continuous rain of which may have enriched
the primitive Earth of carbon usable for synthesising biological material. 

%\subsection{Solar System objects}

%The implications for the chemistry inherited by the 
%Solar System and, in general, by stars formed close to high-mass stars, will be
%discussed. We will focus on the species created during high-mass star formation that can contribute to 
%the chemical enrichment of the environment, with special attention to those important for prebiotic chemistry, 
%such as COMs (e.g.~Beltr\'an et al.~2009), carbon-rich species (e.g.~Fontani et al.~2017), phosphorus-bearing molecules 
%(e.g.~Rivilla et al.~2020), or heavier species containing alkali (Ginsburg et al.~2019) and other rare elements 
%(see McGuire et al.~2018 for a review)

\subsection{Sub-Solar metallicity and extragalactic environments}

Interferometric observations are now able to resolve individual high-mass star-forming regions both in the outer periphery of the Milky Way and in nearby galaxies.
Both environments are characterised by metallicities lower than Solar.
Understanding chemical evolution in star formation at sub-Solar metallicity can give insights into chemical processes at work in the early Galaxy or in
high-redshift galaxies, where the gas was not enriched in metals yet by stellar nucleosynthesis.
Although evolutionary studies have not been performed either in outer Galaxy or in  extragalactic high-mass star forming regions, observational works have highlighted an interesting chemical similarity with regions in the immediate vicinity of the Sun.

In the outer Galaxy the expected scenario would
be that the formation and survival of molecules is less efficient than in regions close to the Sun, owing to both a lower abundance of metals and less shielding from UV radiation.
However, a hot core, WB89-789, associated with emission of COMs (such as CH$_3$OH, CH$_3$CCH, CH$_3$OCHO, and CH$_3$OCH$_3$), was detected at $\sim 19$~kpc from the Galactic centre (Shimonishi et al. 2021), where metallicity is estimated to be a factor of 4 lower than the Solar value according to the Galactic oxygen
abundance gradient \citep[e.g.,][]{arellano21}.
Moreover, several studies indicate that the environmental metallicity does not affect the formation efficiency of most of the
molecules studied so far, but it appears to act mostly as a scaling factor for the molecular abundances relative to the elemental
ones \citep{bernal21,fontani22,gigli25}.
However, the targets observed in the mentioned works all harbour protostellar massive objects whose relative evolutionary stages have not been determined precisely yet.
Significant progresses in this direction can be made in the future by observing the centimetre-continuum emission at high-angular resolution to distinguish between HMPOs, HCHIIs and UCHIIs.
In this vein, the unprecedented sensitivity observations that can be performed in the near future with the Square Kilometre Array (SKA) and the next generation Very Large Array (ngVLA) will be of paramount importance.

In line with these findings, there are now several examples of extragalactic hot cores in the magellanic clouds: \citep{sewilo18,sewilo22,shimonishi16,shimonishi23}.
Considering that the Large Magellanic Cloud (LMC) and Small Magellanic Cloud (SMC) have metallicities smaller than Solar (by a factor 0.5 $Z_{\odot}$ and 0.2 $Z_{\odot}$, respectively), these findings indicate that the formation of HMCs are common phases of the high-mass star formation process also in low metallicity environments.
Interestingly, these extragalactic HMCs show molecular abundances that in some species, like SO$_2$, are consistent among them considering only a metallicity scaling factor, while others, like \METH, are not consistent.
Such difference suggests that the abundance of \METH, and in general organic molecules \citep{shimonishi23}, is influenced less by metallicity than by other physical parameters (e.g. visual extinction, cosmic-ray ionisation rate, UV illumination, etc.).

Moving further away from the Milky Way, \citet{sutter24} have observed with JWST a sample of 19 nearby galaxies from the Physics at High Angular Resolution in Nearby GalaxieS (PHANGS) survey, focussing on the presence of PAHs on 10-50 parsec-scale clouds. 
The PAH fraction steeply decreases in \HII\ regions, revealing the destruction of these small grains in regions of ionised gas. 
Outside \HII\ regions, the PAHs abundance is constant across the PHANGS sample.

\section{Summary and conclusions}
\label{summary}

\begin{figure}
    \centering
    \includegraphics[width=0.95\linewidth]{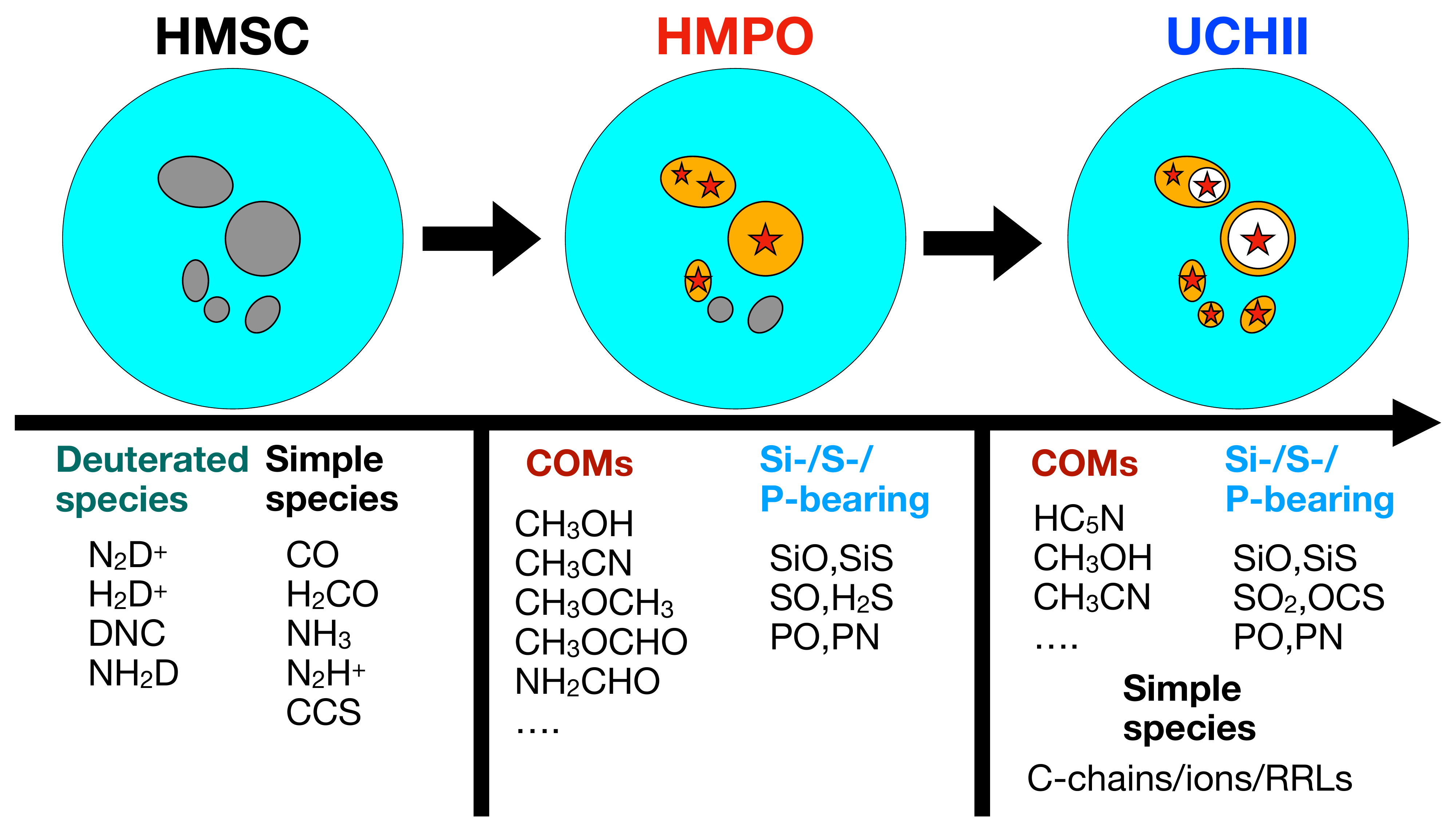}
    \caption{Sketch of the main chemical indicators of each evolutionary stage in the high-mass star formation process.}
    \label{fig:sequence}
\end{figure}

This review article presents a comprehensive overview of recent advances in our understanding of the chemical evolution associated with massive star formation. 
Chemical complexity is emerging as a critical probe of the physical and dynamical processes driving massive star formation, but the intrinsic physical complexity of massive star-forming regions and the short evolutionary timescales make any connection between physics and chemistry challenging.
We summarise key observational results, both from high-resolution interferometric studies and single-dish source surveys, alongside developments in theoretical modelling that address the complex interplay between chemistry, dynamics, and radiative feedback. 
Emphasis is placed mainly on the temporal evolution of molecular complexity, the role of accretion, outflows, and UV photons in shaping chemical signatures, and the constraints that these processes impose on models of high-mass star formation:

\begin{itemize}
\item High-mass star-forming regions host the most chemically-rich sources in galaxies, in particular those sources in the HMPO or the \HC\ and \UC\ evolutionary phase, as indicated by the large number of species detected thanks to unbiased molecular line surveys. The chemistry of these sources is composed mainly of hydrogenated molecules, oxygen-bearing, nitrogen-bearing, sulfur- and silicon-bearing, phosphorus-bearing, and isotopologues and deuterated species, although other species including chlorine, fluorine, sodium, potassium and iron have also been detected toward these massive cores. These regions are also the richest sources of COM emission, including large complex molecules with $>$12 atoms and prebiotic species. 

\item Surveys of high-mass star forming cores divided in evolutionary groups observed in specific species suggest that the best evolutionary indicator is the deuterium fraction computed from \H\ and H$_3^+$.
Deuteration in other species, or other isotopic fractions, seem to be less sensitive to evolution and more to local physical conditions.
\item Tracers whose relationship with evolution was investigated, but for which there is still no general consensus, are: (1) the SO/SO$_2$ abundance ratio, proposed to decrease with evolution; (2) the abundance of COMs, S-, Si-, and P-bearing molecules, proposed to increase with evolution; (3) methanol, water, and OH masers, which are tentatively proposed to appear in this order with time.
\item The decrease with time of \D/\H\ is probably the most convincing evolutionary link between low- and high-mass star forming cores. 
Other ratios such as SO/SO$_2$, and COMs and carbon chain abundances do not seem to behave in the same way. 
\item Chemical modelling of high-mass star-forming regions provides constraints on the timescales of their dynamical evolution. In particular, models of deuterium fractionation in chemical species suggest that the collapse of massive protostars proceeds significantly more slowly than the free-fall rate.
\item Evolutionary sequence of high-mass star-forming regions is not well defined. Although it is evident that gas-phase abundances of complex organic molecules increase along with the development of a protostar from a cold HMSC to a hot core stage, it is not entirely clear at which stage the COMs are formed. 
Recently proposed non-diffusive chemical reactivity opens a possibility to form COMs in amounts observed at hot cores already at the earliest cold stages of high-mass star formation. 
Further studies of deuterated COMs will probably shed light on the dominant mechanism of the formation of COMs.

\end{itemize}

A tentative chemical evolutionary sequence  
is illustrated in Fig.~\ref{fig:sequence}: Simple
abundant molecules and deuterated molecules are the best tracers of the early HMSC phase, while COMs and Si-/S-/P-bearing molecules characterise the HMPO and UCHII stages. 
The main chemical difference between HMPOs and  UCHIIs is an enhanced abundance in UCHIIs of simple carbon-chains and rare ions.
UCHIIs are also characterised by the presence of bright RRLs.

However, from an observational point of view one of the key takeaway messages from this review is that identifying good evolutionary indicators remains a significant challenge.
A major issue arises from the fact that it is still not easy to understand if objects with similar observational properties are in different evolutionary stages. 
In this respect, to establish the elusive presence/absence of embedded indicators of a specific phase is critical. 
For example, an embedded centimetre continuum source can distinguish between a HMPO and a young \HII\ region with a similar molecular envelope.
Similarly, the presence of an embedded young outflow can distinguish between a HMSC and an early HMPO in which the very embedded protostellar object has not significantly affected the collapsing envelope yet.
Another major observational problem mirroring the previous one lies on the fact that source surveys often contain objects almost coeval but with different physical properties (in particular density, temperature, or external irradiation), making it difficult to isolate the role of time in determining the molecular abundances. 

Next generation radio and millimetre interferometers (SKA, ngVLA, the new ALMA receivers) are likely the best facilities to solve these problems, because they can unveil essential source properties, such as faint embedded centimetre continuum sources and/or young outflows, thanks to their unprecedented sensitivity.
They can also perform source surveys of many sources and derive their detailed physical structure in reasonable amounts of observation time.

Finally, a relevant open question is how the chemical composition of ice mantles evolves with time in high-mass star forming regions, and how it is linked to the gas-phase composition and evolution.
JWST observations of ices towards high-mass star-forming cores in well-established evolutionary stages and with well-constrained gas-phase abundances, combined with models specifically tailored for such environments, will be crucial to shed light on this essential and  fascinating open question.

\tiny
 \keyFont{ \section{Keywords:} star formation, interstellar medium, astrochemistry, protostars, } %All article types: you may provide up to 8 keywords; at least 5 are mandatory.

\section*{Conflict of Interest Statement}
%All financial, commercial or other relationships that might be perceived by the academic community as representing a potential conflict of interest must be disclosed. If no such relationship exists, authors will be asked to confirm the following statement: 

The authors declare that the research was conducted in the absence of any commercial or financial relationships that could be construed as a potential conflict of interest.

\section*{Author Contributions}

The authors have totally written this manuscript.

%\section*{Funding}
%Details of all funding sources should be provided, including grant numbers if applicable. Please ensure to add all necessary funding information, as after publication this is no longer possible.

\section*{Acknowledgments}
The work by AV is supported via the project FEUZ-2025-0003.
%The authors are grateful to the reviewers for their constructive comments, which helped to improve the original version of the manuscript.

%\section*{Supplemental Data}
% \href{http://home.frontiersin.org/about/author-guidelines#SupplementaryMaterial}{Supplementary Material} should be uploaded separately on submission, if there are Supplementary Figures, please include the caption in the same file as the figure. LaTeX Supplementary Material templates can be found in the Frontiers LaTeX folder.

%\section*{Data Availability Statement}
%The datasets [GENERATED/analysed] for this study can be found in the [NAME OF REPOSITORY] [LINK].
% Please see the availability of data guidelines for more information, at https://www.frontiersin.org/about/author-guidelines#AvailabilityofData

\bibliographystyle{Frontiers-Harvard}
%  Many Frontiers journals use the Harvard referencing system (Author-date), to find the style and resources for the journal you are submitting to: https://zendesk.frontiersin.org/hc/en-us/articles/360017860337-Frontiers-Reference-Styles-by-Journal. For Humanities and Social Sciences articles please include page numbers in the in-text citations 
%\bibliographystyle{Frontiers-Vancouver} % Many Frontiers journals use the numbered referencing system, to find the style and resources for the journal you are submitting to: https://zendesk.frontiersin.org/hc/en-us/articles/360017860337-Frontiers-Reference-Styles-by-Journal
\bibliography{review_FrASS}

\end{document}